\documentstyle[aps,prd,eqsecnum,twocolumn,epsf]{revtex}

\newcommand{\beq}{\begin{equation}}
\newcommand{\beqn}{\begin{eqnarray}} 
\newcommand{\eeq}{\end{equation}}
\newcommand{\eeqn}{\end{eqnarray}}

\newcommand{\singlefig}[2]{
\begin{center}
\begin{minipage}{#1}
\epsfxsize=#1
\epsffile{#2}
\end{minipage}
\end{center}}
\newenvironment{figcaption}[2]{
 \vspace{0.3cm}
 \refstepcounter{figure}
 \label{#1}
 \begin{center}
 \begin{minipage}{#2}
 \begingroup \small FIG. \thefigure: }{
 \endgroup
 \end{minipage}
 \end{center}}
\def\beq{\begin{equation}}
\def\eeq{\end{equation}}
\newcommand{\gsim}{\mbox{\raisebox{-1.ex}{$\stackrel
     {\textstyle>}{\textstyle\sim}$}}}
\newcommand{\lsim}{\mbox{\raisebox{-1.ex}{$\stackrel
     {\textstyle<}{\textstyle \sim}$}}}
\newcommand{\square}{\kern1pt\vbox{\hrule height
1.2pt\hbox{\vrule width 1.2pt\hskip 3pt
   \vbox{\vskip 6pt}\hskip 3pt\vrule width 0.6pt}\hrule
height 0.6pt}\kern1pt}

\begin{document}
\draft \twocolumn[\hsize\textwidth\columnwidth\hsize\csname
@twocolumnfalse\endcsname

\title{Multi-field fermionic preheating}
\author{Shinji Tsujikawa}
\address{Department of Physics, Waseda University,
3-4-1 Ohkubo, Shinjuku-ku, Tokyo 169-8555, Japan\\[.3em]
Email: shinji@gravity.phys.waseda.ac.jp}
\author{Bruce A. Bassett}
\address{Relativity and Cosmology Group (RCG), University of
Portsmouth, Mercantile House, Portsmouth,  PO1 2EG, 
England\\[.3em]
Email: bruce.bassett@port.ac.uk}
\author{Fermin Viniegra}
\address{Department of Theoretical Physics, Oxford
University, Oxford OX1 3NP, England\\[.3em]
Email: fermin@thphys.ox.ac.uk}
\date{\today}
\maketitle
\begin{abstract}
Fermion creation during preheating in the presence of multiple 
scalar fields exhibits a range of interesting behaviour relevant
to estimating post-inflation gravitino abundances. 
We present  non-perturbative analysis of fermion production
over a $6$-dimensional parameter space in an expanding background  
paying  particular attention to the interplay between 
instant preheating ($\chi$-$\psi$) and direct fermion 
preheating ($\phi$-$\psi$). In the broad resonance regime 
we find that instant fermion production 
is sensitive to suppression of the long wavelength $\chi$ modes 
during inflation. Further, the standard scenario of resonant fermionic 
preheating through  inflaton decay can be significantly modified by 
the $\chi$-$\psi$ coupling, and may even lead to a decrease in the number 
of fermions produced. We explicitly include the  effects of metric 
perturbations and demonstrate that they  are important at small 
coupling but not at strong  coupling,  due to  the rapid 
saturation of the Pauli bound. 
\end{abstract}
\vskip 1pc 
\centerline{WUAP-00/18, RCG-00/17}
\vskip 2pc
]

\section{Introduction}                           %

The evolution of spin $1/2$ (and higher) quantum fields  propagating 
on curved backgrounds has traditionally been considered an 
esoteric domain of little cosmological interest \cite{BD80}. In the face
of the power and simplicity of the inflationary paradigm the added 
complexities associated with their renormalization, regularization
and solution \cite{early,DZ} in the presence of a non-trivial
metric $g_{\mu\nu}$ made them unpopular topics of study.  

The last two years have, however,  
seen a reversal in the fortune of higher-spin fields
with the growing belief that they may be relevant for cosmology in a 
number of rather profound ways.  Decay of the inflaton 
to massive  states {\em during inflation} \cite{BK} can lead to non-trivial 
features in the primordial power spectrum and 
cosmic microwave background (CMB) \cite{CKRT}. In inflationary 
scenarios it is during 
reheating and  preheating -- the non-equilibrium end to 
many inflationary models -- that fermions first became  
important for the subsequent evolution and nature of the universe. 

There are three types of inflaton decay products of significant  
current interest: (i)  gravitinos, the supersymmetric partner of the 
graviton present  in supergravity (SUGRA) \cite{review},  
(ii) heavy right-handed neutrinos which may 
have mediated lepto-genesis \cite{lepto} and (iii) the so-called 
{\em cryptons}\cite{cryptons}  or {\em wimpzillas}\cite{wimps}  --  
massive ($\sim 10^{12}$ GeV) 
dark matter which may  account for the cosmic rays observed 
beyond the Greisen-Zatsepin-Kuzmin (GZK) cut-off. 

While both heavy right-handed neutrinos  
and cryptons provide potential solutions to serious problems, the spin-$3/2$ 
gravitini are a plague. A typical constraint is that their number 
density $n_{3/2}$ must satisfy the 
bound $n_{3/2}/s~\lsim~10^{-14}$ where $s$ is the entropy density of the 
universe at reheating for a gravitino of 
mass $m_{3/2} \sim {\cal O}(100)$ GeV.

In a perturbative theory of reheating\cite{oldre}, this leads to the 
constraint 
$T_r~\lsim~10^8$ GeV\cite{subir2,review}, on the reheat temperature, $T_r$. 
Early studies of gravitino production\cite{earlypre} 
following preheating\cite{pre1,pre2,KLS97,BDHS}   
argued that thermal gravitino abundances were significantly 
enhanced due to  the non-perturbative nature of preheating. Nevertheless, 
it was only recently realized that cosmological bounds arising from 
preheating are much tighter than previously  
imagined\cite{MM,gpro1,gpro2,lemoine,MP99,gpro3}.

In these latter works the direct production of gravitini was studied through 
their coupling to the inflaton, $\phi$. 
Nevertheless, in a general SUGRA theory 
one expects many scalar fields $\phi_i$ to contribute to 
the  K\"ahler potential $K(\phi_i\phi^i{}^*)$. Since the  gravitino 
effective mass is given by \cite{gpro1}
\beq
m_{3/2}(\phi_i) = e^{K/2} \frac{W}{m_{\rm pl}^2}
\label{gravmass}
\eeq
where $W$ is the superpotential and $m_{\rm pl} \sim 10^{19}$ GeV is
the Planck mass,  it  is natural to ask how other scalar fields will 
affect gravitino production. 

Since we are interested in general results valid also for 
leptogenesis scenarios   we will not work directly with the 
gravitino (Rarita-Schwinger) equations 
but rather with the spin $1/2$ Dirac equation\cite{DK}. Fortunately this
gives useful results for gravitino production due to the nature of the 
helicity-$3/2$ and $1/2$ evolution equations.
On the one hand the helicity-$3/2$ mode reduces directly to a 
Dirac equation \cite{DZ,MM}  and on the other, 
the dangerous helicity-$1/2$ mode,
which is not suppressed in the $M_{\rm pl} \rightarrow \infty$ limit 
\cite{gpro1}, behaves like the goldstino in the high momentum 
limit \cite{GRT,MP99,gpro3}.

We will consider a rather general potential leading to a  fermion 
effective mass 
\beq
m_{\rm eff} = m_{\psi} + h_1 \phi + h_2 \chi\,,
\label{effmass}
\eeq
where $m_{\psi}$ is the bare mass of fermion, $\phi$
and $\chi$ are the inflaton and a scalar field coupled to inflaton,
respectively. 
The addition of the $h_2$ coupling modifies the standard production in two 
interesting cases (i) if $\langle \chi \rangle \neq 0$ during inflation  
or  (ii) if $\chi$ itself undergoes 
resonant  amplification during preheating. We will consider both 
cases.

While we will draw general conclusions regarding multi-field 
resonant fermion production there are certain issues we 
will not address. From Eq. (\ref{effmass}) it is clear 
that  in general the fermion effective mass will have more than one 
natural frequency, analogous to the situation with scalar fields.
If the natural frequencies (of $\phi$ and $\chi$) are irrationally 
related then the effective mass is quasi-periodic\cite{BB98}. 
In the scalar case this causes a dramatic increase in the strength and 
breadth of the resonance bands. 

Similarly, if the fermion 
is coupled to many scalar fields the 
effective mass (\ref{effmass}) may become chaotic or stochastic. In the 
scalar case this is known to again enhance the resonance if the 
time correlations are small\cite{BT98}. Whether these effects  
occur in the fermion case too is unknown at present. 

Instead in this work we will investigate in detail the following issues:

$\bullet$ What is the interplay between direct (through the coupling 
$h_1$) and instant (through $h_2$) \cite{instant} 
fermion  preheating ? 

$\bullet$ How is instant preheating sensitive to the $\chi$ vacuum 
expectation value (VEV) during inflation ? This will strongly 
alter the fermion  effective mass, Eq. (\ref{effmass}), and 
therefore the nature of  stochastic fermion production.

$\bullet$ How does the inclusion of metric perturbations affect fermion
production ? Are their effects similar to the scalar case ? 

To answer these issues requires a $6$-dimensional parameter space 
(illustrated in Fig.~\ref{Fig1}). We then study the 
evolution of the spectrum of 
massive spin-$1/2$  fermions over this ``moduli'' space of 
fermionic preheating.  

In Sec.~\ref{model} we discuss this parameter space and 
present the equations for the background geometry and perturbations. 
Fermionic preheating  is then analysed in the massless 
(Sec.~\ref{massless}) and massive cases  (Sec.~\ref{massive}).

\begin{figure}
\begin{center}
\singlefig{7cm}{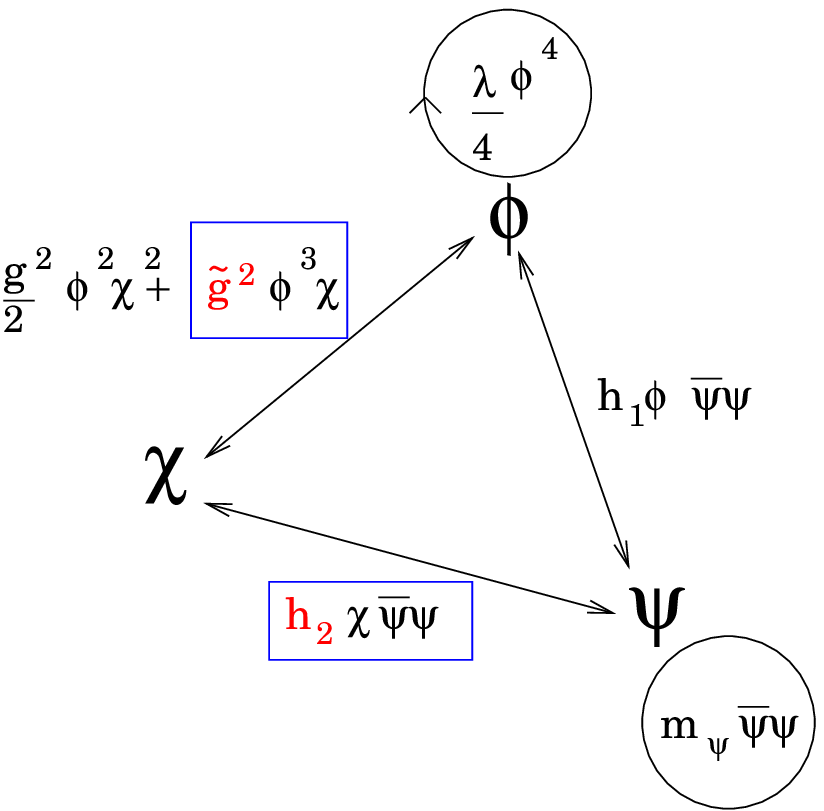}
\begin{figcaption}{Fig1}{8cm}
Schematic illustration of the potential (\ref{B12}). We consider the 
inflaton $\phi$, scalar field $\chi$ and fermion  $\psi$. 
Interactions are shown as lines with arrows, leading to a 
$6$-dimensional parameter space of couplings and masses. The new couplings 
we consider in this paper are shown in red and are boxed. 
\end{figcaption}
\end{center}
\label{fig1}
\end{figure}

\section{The model and basic equations}\label{model}

Let us consider the massless chaotic inflation model
\begin{eqnarray}
V &=& \frac14 \lambda \phi^4 + \frac12 g^2\phi^2\chi^2 + 
\tilde{g}^2 \phi^3 \chi \nonumber \\ 
&+&  (m_{\psi}+h_1\phi+h_2\chi) \overline{\psi}\psi,
\label{B12}
\end{eqnarray}
where $\psi$ is the fermion field of interest with bare mass 
$m_{\psi}$, which is coupled to the inflaton, $\phi$,  and a scalar field 
$\chi$ through the couplings  $h_1$ and $h_2$ respectively. 
Typically we adopt the value $\lambda=10^{-12}$ for the self-coupling 
which comes from  CMB constraints.

The perturbative (tree-level) decay rates 
for the inflaton and the $\chi$ field to fermions are\cite{oldre} 
\beqn
\Gamma(\phi \rightarrow \psi\psi) &=& \frac{h_1^2m_{\phi}}{8\pi} = 
\frac{h_1^2}{8\pi}(\lambda \phi^2 + g^2\chi^2+3\tilde{g}^2\phi\chi )^{1/2}
\nonumber \\ 
\Gamma(\chi \rightarrow \psi\psi) &=& 
\frac{h_2^2m_{\chi}}{8\pi} = \frac{h_2^2 g|\phi|}{8\pi}
\eeqn
However, these decay rates   
cease to be accurate during resonance and after 
a fermion-scalar plasma has formed. Boyanovsky {\em et al.}\cite{BD1,BD2}, 
Baacke {\em et al.}\cite{BHP,BP} and Ramsey {\em et al.}\cite{RHS}
studied this resonant, non-equilibrium 
production of fermions in the absence of  the $\chi$ field. 

Greene \& Kofman\cite{GK} gave 
analytical insight into resonant fermion production and recent 
work has brought out the strong analogy with the scalar 
stochastic  resonance regime \cite{GPRT,bellido,GK2,PS}.
We extend these works to the multi scalar field case where 
the $\chi$ field is coupled to both of $\psi$ and $\phi$ fields.

We have also included the $\tilde{g}^2 \phi^3 \chi$ coupling
which may for example, appear in  SUGRA models\cite{SUGRA} in addition 
to the usual  $\frac12 g^2\phi^2\chi^2$ coupling.
Due to the existence of this term, $\langle \chi \rangle \neq 0$ during 
inflation and the exponential suppression 
for the $\chi$ field during inflation is avoided \cite{mpre5,shinji}. 
We will see later that $\tilde{g}$ can significantly enhance the
$\chi$-$\psi$ decays.

When $m_{\psi} = 0$, the complete system of 
equations is  conformally invariant (modulo the small  conformal couplings
needed for $\phi$ and $\chi$).
Conformally rescaled  fields will be capped with a $~{}\tilde{}$, i.e. 
$\tilde{\phi} \equiv a \phi$, $\tilde{\chi} \equiv a\chi$,
and $\tilde{\psi} \equiv a^{3/2} \psi$ with $a$ the scale factor.

\subsection{The background and scalar field equations}
In this paper, we work around a flat 
Friedmann-Lema$\hat{\i}$tre-Robertson-Walker (FLRW) background
with  perturbations in the longitudinal gauge
\begin{eqnarray}
ds^2=-(1+2\Phi)dt^2
+a^2(1-2\Psi)\delta_{ij} dx^i dx^j,
\label{B1}
\end{eqnarray}
where $\Phi$ and $\Psi$ are gauge-invariant potentials\cite{MFB,mpre2b}.
The relation  $\Phi=\Psi$ follows since the anisotropic stress vanishes 
at linear order \cite{mpre2b}.
Decomposing the scalar fields into homogeneous parts and fluctuations
 $\phi(t, {\bf x}) \to \phi(t)+\delta\phi(t,{\bf x})$, 
$\chi(t, {\bf x}) \to \chi(t)+\delta\chi(t,{\bf x})$, 
one obtains the equations for the Fourier modes of the scalar field  
perturbations:
\begin{eqnarray}
& &\delta\ddot{\phi}_k + 3H\delta\dot{\phi}_k+
\left(\frac{k^2}{a^2}+3\lambda \phi^2
+g^2\chi^2+6\tilde{g}^2\phi\chi \right)\delta\phi_k
\nonumber \\
&=& 4\dot{\phi} \dot{\Phi}_k 
+ 2(\ddot{\phi}
+3H\dot{\phi})\Phi_k-(2g^2\phi\chi+3\tilde{g}^2\phi^2)
\delta\chi_k, \nonumber \\
\label{B16}
\end{eqnarray}
\begin{eqnarray}
\delta\ddot{\chi}_k &+& 3H\delta\dot{\chi}_k+
\left( \frac{k^2}{a^2}+g^2\phi^2 \right) \delta\chi_k
= 4\dot{\chi} \dot{\Phi}_k \nonumber \\
&+& 2(\ddot{\chi}+3H\dot{\chi})\Phi_k
-(2g^2\phi\chi+3\tilde{g}^2\phi^2)\delta\phi_k. 
\label{B17}
\end{eqnarray}
A convenient equation for the potential  $\Phi_k$ is 
\begin{eqnarray}
\dot{\Phi}_k+H\Phi_k=4\pi G
(\dot{\phi} \delta\phi_k+\dot{\chi} \delta\chi_k),
\label{B18}
\end{eqnarray}
where $G \equiv m_{\rm pl}^{-2}$ is Newton's 
gravitational constant. 
In Eq.~$(\ref{B18})$, we neglect the contribution
of fermions since it is second order and typically negligible 
relative to the  scalar fields due to the Pauli exclusion principle. 

We include the second order energy densities in the scalar field  
fluctuations in the evolution
of the background quantities via $k$-space integrals.  
The equations for the Hubble parameter ($H \equiv \dot{a}/a$)  
and homogeneous components of the 
scalar fields are thus given by \cite{structure}
\begin{eqnarray}
 H^2 &=&
  \frac{8\pi G}{3}
     \Biggl[ \frac12 \dot{\phi}^2+ \frac12
    \langle \delta \dot{\phi}^2 \rangle+
     \frac{1}{2a^2} \langle (\nabla \delta\phi)^2 
      \rangle  \nonumber \\
     &+& \frac14 \lambda(\phi^4+6\phi^2
      \langle \delta\phi^2 \rangle
    +3\langle \delta\phi^2 \rangle^2) 
     +\frac12 \dot{\chi}^2 
    +\frac12  \langle \delta \dot{\chi}^2 \rangle \nonumber \\
    &+& \frac{1}{2a^2} \langle (\nabla \delta\chi)^2  \rangle
     +\frac{g^2}{2}\langle \phi^2\chi^2 \rangle
     +\tilde{g}^2 \langle \phi^3\chi \rangle
\Biggr],
\label{B13}
\end{eqnarray}
\begin{eqnarray}
\ddot{\phi} &+& 3H \dot{\phi} +\lambda \phi
(\phi^2+3 \langle \delta\phi^2 \rangle)  \nonumber \\
&+& g^2(\chi^2+\langle \delta\chi^2 \rangle) \phi
+3\tilde{g}^2(\phi^2+\langle \delta\phi^2 \rangle)\chi=0,
\label{B14}
\end{eqnarray}
\begin{eqnarray}
\ddot{\chi} &+& 3H \dot{\chi} +g^2
(\phi^2+\langle \delta\phi^2 \rangle)\chi \nonumber \\
&+& \tilde{g}^2(\phi^3+3\phi\langle \delta\phi^2 \rangle)=0,
\label{B15}
\end{eqnarray}
where the expectation values of $\delta \phi^2$ and 
$\delta\chi^2$ are defined as 
\begin{eqnarray}
\langle \delta \phi^2 \rangle = \frac1{2\pi^2} \int k^2
|\delta \phi_k|^2 dk\,,
\end{eqnarray}
\begin{eqnarray}
\langle \delta\chi^2 \rangle =\frac1{2\pi^2} \int k^2
|\delta \chi_k|^2 dk\,.
\end{eqnarray}
As for other definitions of the fluctuational quantities, see e.g., 
Ref.~\cite{Boy}.  
In the present model, both  $\langle \delta \phi^2 \rangle$
and $\langle \delta\chi^2 \rangle$ grow by parametric resonance 
as shown in Fig. (5), and alter  the evolution of the background fields. 

We implement backreaction using the Hartree approximation (i.e. 
neglecting rescattering effects \cite{KLS97}) and ignore
the backreaction effect due to  the production of fermions which 
is typically justified because of the small amount of energy in the fermions 
by the exclusion principle (very massive fermion production being an 
obvious example where this is not necessarily true).
The Hartree approximation alters the terms $\phi^2$ and $\chi^2$ to 
$\langle\phi^2 \rangle \equiv \phi^2+\langle \delta \phi^2 \rangle$ and
$\langle\chi^2 \rangle \equiv \chi^2+\langle \delta \chi^2 \rangle$ 
in Eqs.~$(\ref{B16})$, $(\ref{B17})$, respectively.
The variances are crucial to ensure conservation of energy.

\subsection{The Dirac equation}

The Dirac equation in curved spacetime is given in general by\cite{BD80}
\begin{eqnarray}
i\gamma^{\mu} \left(\partial_\mu+\frac{1}{8}
\left[\gamma^b,\gamma^c\right] e_b^\nu e_{c \nu;\mu}\right) \psi
-m_{\rm eff} \psi=0,
\label{B2}
\end{eqnarray}
where the Latin indices $a,b,c,$ and Greek indices $\mu,\nu$
refer to the locally inertial (tetrad) coordinates and 
the general curvilinear (spacetime) coordinates respectively.
The vierbeins $e_a^\mu$ connect these coordinate systems via
\beq
ds^2 = g_{\mu \nu} dx^{\mu} dx^{\nu} = \eta_{ab} e^a e^b,
\eeq
where $g_{\mu \nu}$ and $\eta_{ab}$ are the curved space and Minkowskian 
metrics respectively. The $\gamma^{\mu}$ are the curved-space Dirac 
matrices satisfying the Clifford algebra anti-commutator relation
$\{\gamma^{\mu},\gamma^{\nu}\} = 2g^{\mu \nu}$.

Since the spinor $\psi$ has no homogeneous component, 
writing the general Dirac equation to first order  
in  the perturbed metric (\ref{B1})
does not yield any terms containing $\Phi_k$ [in contrast with 
the scalar case, Eqs.~(\ref{B16}), (\ref{B17})], and simply gives
\begin{equation}
(i\gamma^{\mu}\partial_{\mu} -m_{\rm eff}a) \tilde{\psi}=0\,,
\label{B3}
\end{equation}
where $\partial_0$ denotes the derivative with respect to 
conformal time $\eta=\int a^{-1}dt$.
We decompose the $\tilde{\psi}$ field into Fourier
components as
\begin{eqnarray}
\tilde{\psi}=  \frac{1}{(2\pi)^{3/2}} \int d^3 k
&\sum_{s}& [a_s(k) \tilde{{\bf u}}_s(k,\eta)
e^{+i {\bf k} \cdot {\bf x}} \\ \nonumber 
&+&
b_s^{\dagger}(k) 
\tilde{{\bf v}}_s (k,\eta)e^{-i {\bf k} \cdot {\bf x}} ]. 
\label{B25}
\end{eqnarray}
Imposing  the following standard ansatz\cite{GK}:
\begin{eqnarray}
\tilde{{\bf u}}_s(k,\eta)
=(-i\gamma^{\mu}\partial_{\mu}
-m_{\rm eff}a) \tilde{\psi}_k(t) R_{\pm}(k),
\label{B10}
\end{eqnarray}
where $R_{\pm}(k)$ are the eigenvectors of the helicity operator,
which satisfy the relation 
$\gamma^0 R_{\pm}(k)=1$ and ${\bf k} \cdot \sum
R_{\pm}(k)= \pm 1$,
we obtain the mode equation for the $\tilde{\psi}_k$: 
\begin{eqnarray}
\left[ \frac{d^2}{d\eta^2}+k^2+(m_{\rm eff}a)^2
-i\frac{d}{d\eta}(m_{\rm eff}a) \right]
\tilde{\psi}_k = 0.
\label{B45}
\end{eqnarray}
Introducing dimensionless quantities 
\begin{equation}
f \equiv \frac{\tilde{m}_{\rm eff}}{\sqrt{\lambda}\phi(0)},~~~
\kappa^2 \equiv \frac{k^2}{\lambda \phi^2(0)},~~~
x \equiv \sqrt{\lambda}\phi(0) \eta,
\label{f}
\end{equation}
with 
\beqn
\tilde{m}_{\rm eff} &\equiv& a m_{\rm eff}\\ \nonumber 
&=&  a m_{\psi} + h_1 \tilde{\phi} +  h_2 \tilde{\chi}\,,
\eeqn
Eq.~(\ref{B45}) becomes
\begin{eqnarray}
\left[ \frac{d^2}{dx^2}+\kappa^2+f^2-i \frac{df}{dx} \right]
\tilde{\psi}_k = 0.
\label{B20}
\end{eqnarray}
which has a WKB-form solution given  by
\begin{eqnarray}
\tilde{\psi}_k=\alpha_k N_+e^{-i \int_0^t \Omega_k dx}
+\beta_k N_- e^{+i \int_0^t \Omega_k dx},
\label{B21}
\end{eqnarray}
where $\Omega_k^2 \equiv \kappa^2+f^2$ and 
$N_{\pm} \equiv 1/\sqrt{2\Omega_k(\Omega_k \pm f)}$. 
The comoving number density of produced fermions is 
given in terms of the  Bogoliubov coefficients by
\begin{eqnarray}
n_k &\equiv& |\beta_k|^2  \\ \nonumber 
&=&\frac12 -\frac{\kappa^2}{\Omega_k} 
{\rm Im} \left(\tilde{\psi}_k \frac{d \tilde{\psi}_k^*}{dx} \right)
-\frac{f}{2\Omega_k}.
\label{B22}
\end{eqnarray}
The initial conditions are chosen to be $\alpha_k(0)=1$, $\beta_k(0)=0$,
which corresponds to $n_k(0)=0$. The Bogoliubov coefficients
satisfy the relation $|\alpha_k(t)|^2 + |\beta_k(t)|^2 = 1$, which means
that the exclusion principle restricts the number density of fermions
to below unity, $n_k(t) \le 1$. In the subsequent 
sections, we investigate the dynamics of 
fermionic preheating taking into account both the $h_1$
and $h_2$ coupling.

\section{Massless fermion production}\label{massless}

It was pointed out by several authors\cite{BHP,GK} that {\it massless} 
fermions are resonantly produced by the interaction between the coherently 
oscillating inflaton  and fermion field during preheating. The 
case of most interest $q_1 \equiv h^2_1/\lambda \gg 1$ leads
to strongly non-adiabatic evolution $(\dot{\Omega}_k/\Omega_k^2 \gg 1)$ 
of the $\tilde{\psi}_k$ only for the short periods when $\tilde{m}_{\rm eff} 
\simeq 0$. For $h_2 = 0$, this coincides with $\phi = 0$ and leads to 
efficient fermion production for momenta 
$\kappa~\lsim~q_1^{1/3}$. 

In most previous works of fermion production, the interaction between 
$\psi$ and $\chi$  fields has been neglected.
However, since the $\chi$ field can be amplified during preheating,
it is expected that this will lead to the fermion production even in the 
absence of the direct ($h_1$) coupling. 

The strength of the $\chi$ field resonance depends on  $g$.
As was analytically  studied in \cite{structure}, the strongest resonance
occurs around $\kappa=0$ when $g^2/\lambda =  2l^2$ where
$l$ is an integer. Hence $\chi$ particle production is maximally efficient 
even if $g^2/\lambda={\cal O}(1)$, which is different from the massive
inflation model where efficient preheating  typically requires large 
couplings  $g~\gsim~3 \times 10^{-4}$. 

\subsection{ light $\chi$ field: $g^2/\lambda = {\cal O}(1)$}

Let us first consider fermions with vanishing bare mass $(m_{\psi} = 0$)  
with $g^2/\lambda= 2$ and $\tilde{g}=0$, which leads to largest growth
rate (Floquet index) for the super-Hubble $\kappa \rightarrow 0$ modes.
We have numerically solved the background and perturbed equations 
and in Fig.~\ref{Fig2} the evolution of 
the comoving number density of fermions.   Plotted are the cases 
of $q_2\equiv h_2^2/\lambda=1, 10, 100$ and $q_1=0$  
for the mode $\kappa=1$.
We start integrating from the end of inflation, 
and take the initial scalar field values to $\phi(0)=0.5m_{\rm pl}$
and $\chi(0)=10^{-9}m_{\rm pl}$.
Since the $\chi$ field grows exponentially through parametric 
resonance, this changes the frequency of the fermion field even in 
the absence of the $h_1$ coupling.

In fact, $n_k$ increases stochastically until
$x \approx 50$ due to the rapid growth of the $\chi$ field.
Not surprisingly we find in Fig.~\ref{Fig2} that 
increased coupling $h_2$ leads to 
larger occupation numbers. For couplings  $q_2~\gsim~100$, 
the final number density reaches of order unity and further increases  
($q_2 \gg 100$)  cause $n_k$ to increase more rapidly.

\begin{figure}
\begin{center}
\singlefig{9cm}{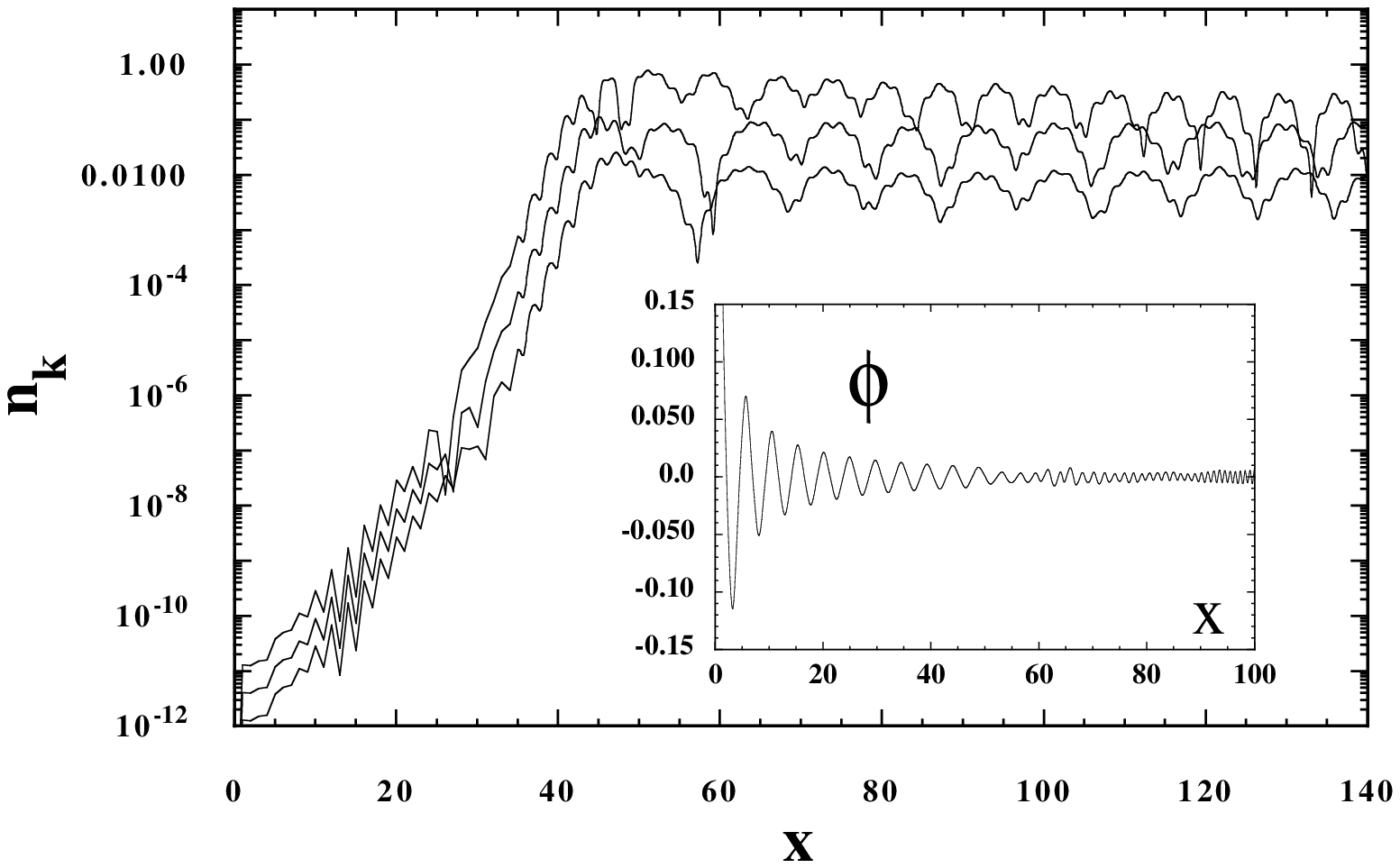}
\begin{figcaption}{Fig2}{9cm}
The evolution of the comoving number density of fermions 
$n_{k}$ versus dimensionless conformal time $x$ during preheating 
in the case of $q_2=1, 10, 100$ (bottom, middle, top)
and $q_1=0$ with $g^2/\lambda=2$ and $\tilde{g}=0$
for the mode $\kappa=1$.
The number density grows exponentially until the growth of the $\chi$
fluctuation stops by backreaction effects.   
We choose the initial values of scalar fields to be $\phi(0)=0.5m_{\rm pl}$
and $\chi(0)=10^{-9}m_{\rm pl}$, which we use in all figures in which 
$\tilde{g} = 0$.{\bf Inset}: The evolution of the inflaton condensate, 
$\phi$, vs $x$. The backreation effect becomes relevant for $x~\gsim~50$.
\end{figcaption}
\end{center}
\end{figure}

\begin{figure}
\begin{center}
\singlefig{9cm}{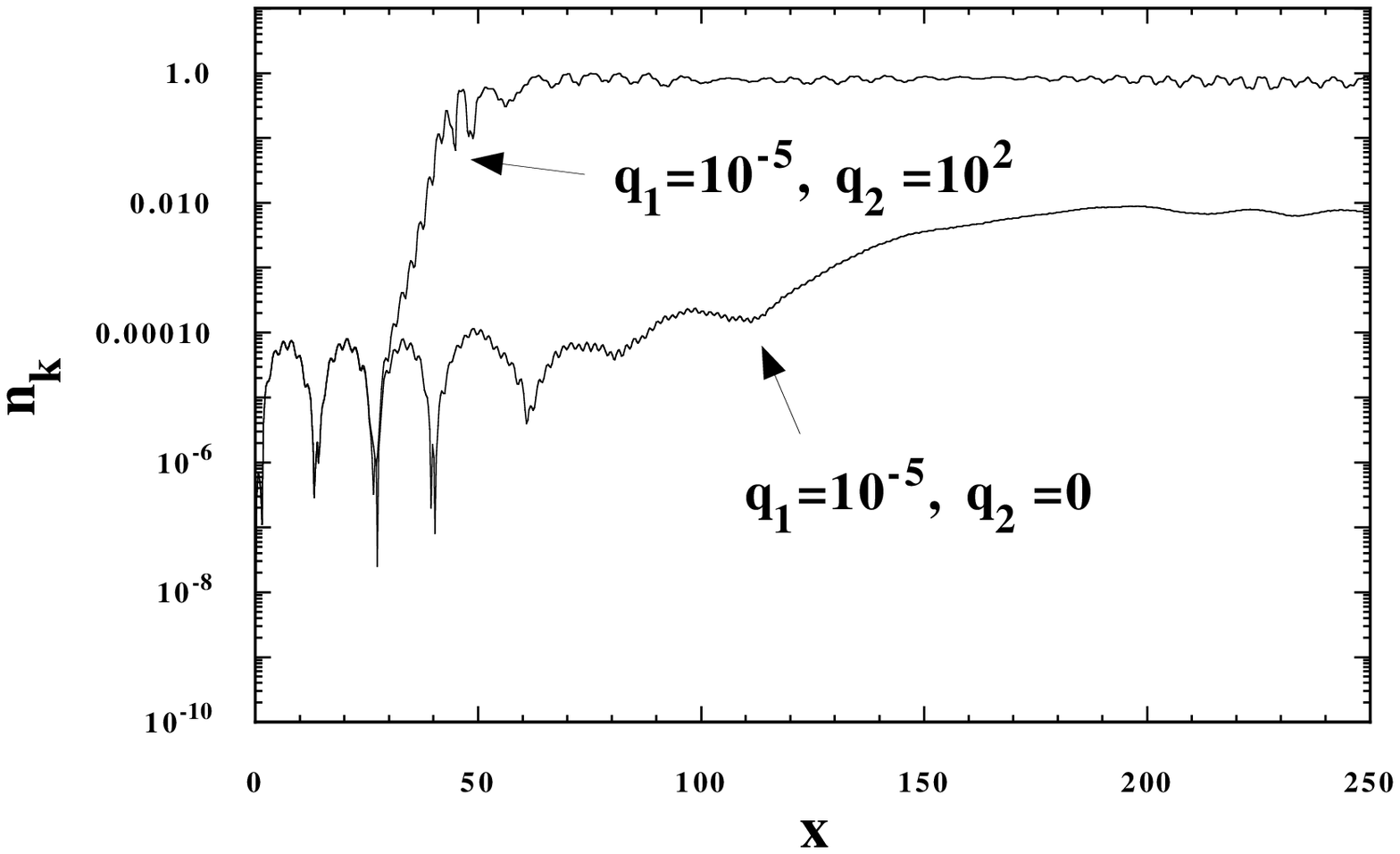}
\begin{figcaption}{Fig3}{9cm}
The evolution of the comoving number density of fermions 
$n_k$ during preheating in the cases of $q_1=10^{-5}, 
q_2=10^2$ (top) and $q_1=10^{-5}, q_2=0$ (bottom) with 
$g^2/\lambda=2$ and $\tilde{g}=0$ for the mode $\kappa=1$.
The initial values of the scalar fields are $\phi(0)=0.5m_{\rm pl}$
and $\chi(0)=10^{-9}m_{\rm pl}$.
The existence of the $h_2$ coupling clearly assists fermion production in
the later phases of preheating once the $\chi$ field has grown significantly 
through resonance.
\end{figcaption}
\end{center}
\end{figure}
For small momentum modes $\kappa \sim 0$, the excitation of fermions 
is stronger than the sub-Hubble mode in Fig.~\ref{Fig2} because the 
nonadiabatic change of the fermion frequency is more relevant.
After  $\chi$ particle production stops by backreaction effects,
the fermion production also terminates, but does not oscillate. 

We should mention that the final occupation number of fermions 
depends not only on $h_2$ but also on the value of $\chi$ 
at the onset of preheating. In the case of $g^2/\lambda={\cal O}(1)$,
since the $g\phi$ term in the lhs of Eq.~$(\ref{B15})$
is comparable to the Hubble rate, the $\chi$ field is hardly damped
during inflation\cite{mpre4}. Moreover, even for an initial value of 
$\chi(0)=10^{-9}m_{\rm pl}$, which is much smaller than $\phi(0)$,
fermions can be created purely by the $h_2$ coupling.

When the interaction, mediated by $h_1$, between inflaton 
and fermion is taken into account,
fermion production at the initial stage is mainly dominated  
by the $h_1$ coupling unless $h_2 \gg h_1$ 
because $\chi$ is so much  smaller than $\phi$
at the beginning of preheating. However, parametric amplification
of the $\chi$ field leads to an additional growth of 
the occupation number of fermions.

In Fig.~\ref{Fig3}, we plot the evolution of $n_k$
for the cases of $q_1=10^{-5}, q_2=10^2$ and  
$q_1=10^{-5}, q_2=0$ with $g^2/\lambda=2$ for the mode $\kappa=1$.
We find that the $h_2$ coupling assists fermion production 
for $x~\gsim~30$, leading to the final density $n_k \sim 1$.
In Fig.~\ref{Fig4}, the spectrum of the final number density $n_k$ is depicted.
When $q_1=10^{-5}$ and $q_2=0$, $n_k$ does not reach 
of order unity except for momenta close to $\kappa=0$. 

\begin{figure}
\begin{center}
\singlefig{9cm}{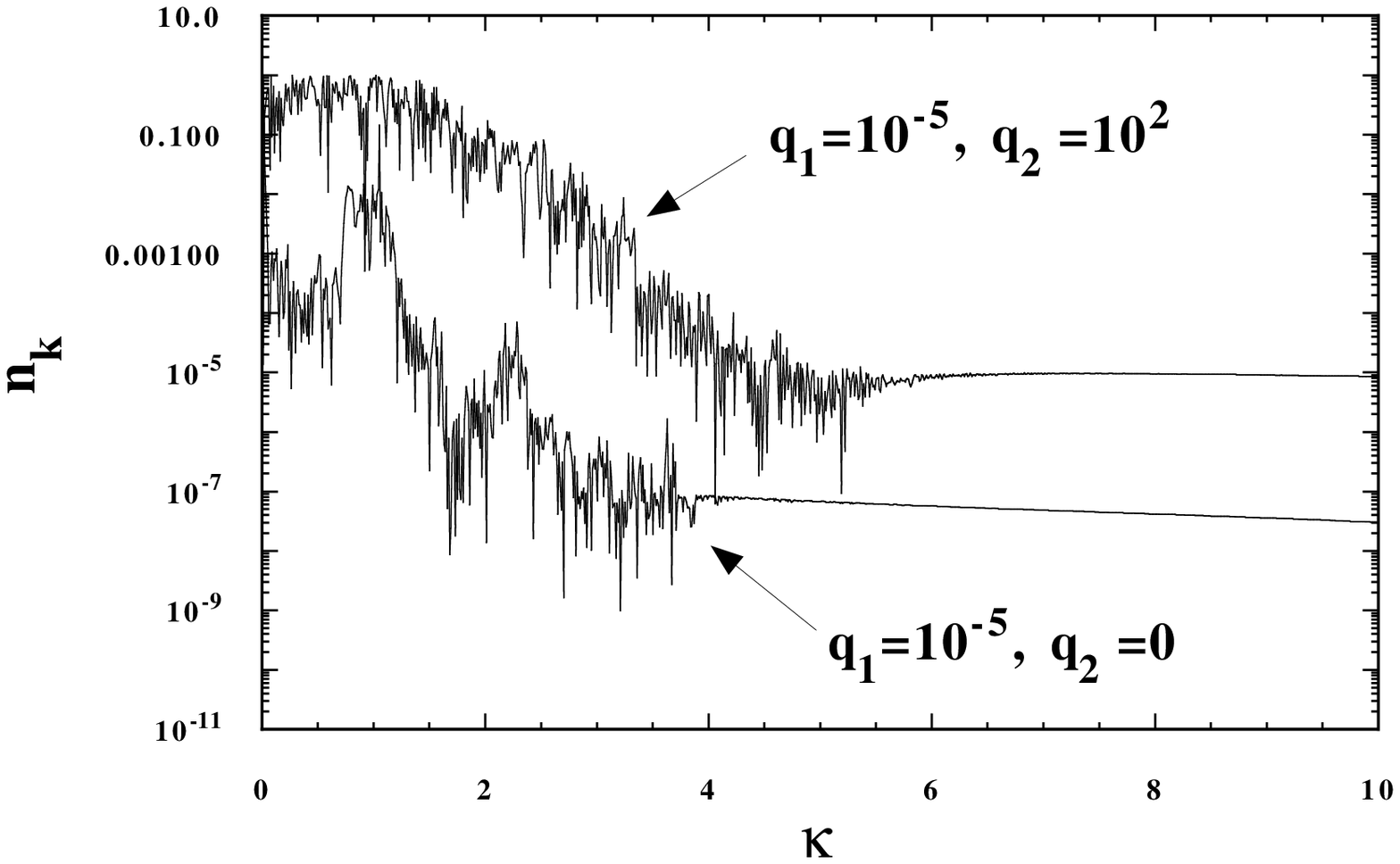}
\begin{figcaption}{Fig4}{9cm}
The spectra of the final comoving number density of fermions $n_k$ 
in the cases of $q_1=10^{-5}, q_2=10^2$ (top) and  
$q_1=10^{-5}, q_2=0$ (bottom) with $g^2/\lambda=2$ and $\tilde{g}=0$.
The initial values of scalar fields are chosen as $\phi(0)=0.5m_{\rm pl}$
and $\chi(0)=10^{-9}m_{\rm pl}$.
\end{figcaption}
\end{center}
\end{figure}

In the case of $g^2/\lambda={\cal O}(1)$, $\chi$ and super-Hubble
$\delta\chi_k$ modes are not damped during inflation\cite{mpre4} 
and have a nearly flat spectrum, as opposed to a $k^3$ spectrum  
when $g^2/\lambda \gg 1$ \cite{suppression,suppression2}. 

This means that when the system enters  preheating, the growth of 
$\chi$ and $\delta\chi_k$ stimulate the growth of super-Hubble $\Phi_k$ 
modes as can be seen in Eq.~$(\ref{B18})$  and in  Fig.~\ref{Fig5}.
The growth rate of the $\delta\chi_k$ fluctuation for the low momentum
mode $\kappa \sim 0$ is large as $\mu=0.2377$ 
for $g^2/\lambda=2$\cite{structure}.

In the standard adiabatic inflation scenario, the Bardeen parameter 
\begin{eqnarray}
\zeta_k \equiv \frac{H(\dot{\phi}Q_k^{\phi}+\dot{\chi}Q_k^{\chi})}
{\dot{\phi}^2+\dot{\chi}^2},
\label{B43}
\end{eqnarray}
is time-independent in the long-wavelength limit, a result which 
simply follows from energy conservation \cite{long}. Here 
$Q_k^{\phi} \equiv \delta \phi_k+\dot{\phi}\Phi_k/H$ and 
$Q_k^{\chi} \equiv \delta \chi_k+\dot{\chi}\Phi_k/H$ are the Sasaki-Mukhanov
variables.
However, in the multi-field case where super-Hubble metric perturbations are
enhanced,  amplification of isocurvature scalar modes leads to amplification
of $\zeta_k$ on large scales {\em before backreaction terminates the 
resonance}  as shown in the inset of Fig.~\ref{Fig5}.

When  $1<g^2/\lambda<3$ where super-Hubble 
modes of the $\delta\chi_k$ fluctuation are enhanced, super-Hubble metric 
perturbations are amplified unless $\chi$ is initially very small.
In the simulation of Fig.~\ref{Fig5} where the initial value of $\chi$ 
is chosen as
$\chi(0)=10^{-9}m_{\rm pl}$, super-Hubble $\Phi_k$ modes increase to the 
order of 0.1. When $\chi(0)~\gsim~10^{-8}m_{\rm pl}$, we numerically 
find that metric perturbations on large scales grow to the nonlinear level
$\Phi_k \sim 1$.

\begin{figure}
\begin{center}
\singlefig{9cm}{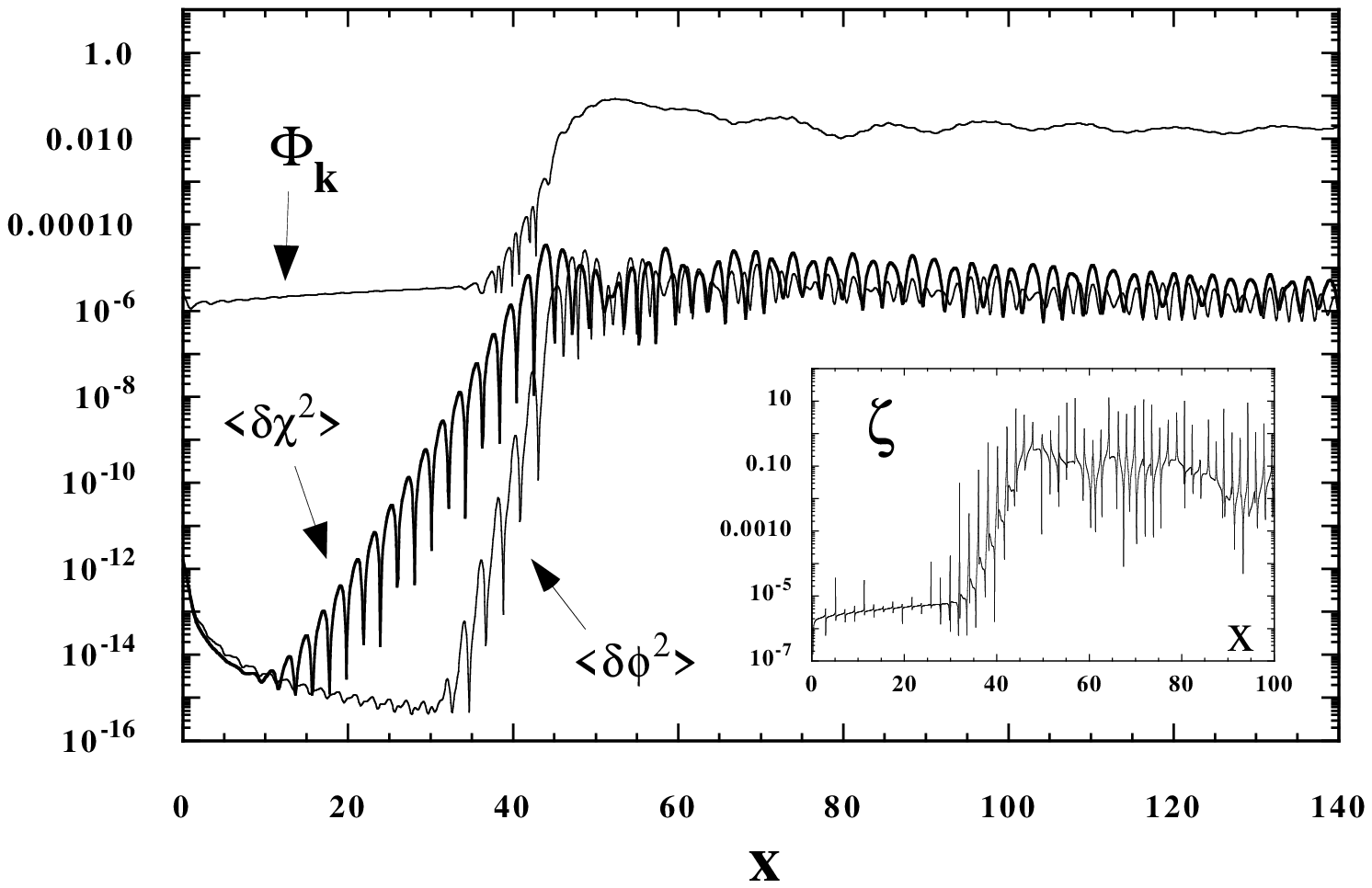}
\begin{figcaption}{Fig5}{9cm}
The evolution of the super-Hubble metric perturbation 
$\Phi_k$ with mode $\kappa=10^{-26}$, and field variances
$\langle\delta\chi^2\rangle$ and 
$\langle\delta\phi^2\rangle$ for $g^2/\lambda=2$ and $\tilde{g}=0$.
We choose the initial values of scalar fields as
$\phi(0)=0.5m_{\rm pl}$ and $\chi(0)=10^{-9}m_{\rm pl}$.
When the $\delta\chi_k$ fluctuation is sufficiently amplified, both 
$\Phi_k$ and $\delta\chi_k$ begin to grow. Finally, backreaction effects
of produced particles shut off the resonance.
{\bf Inset}: The evolution of the Bardeen parameter $\zeta_k$
for the super-Hubble mode $\kappa=10^{-26}$. The growth of $\zeta_k$ is 
due to the fact that the $\chi$ field is light during inflation for 
$g^2/\lambda = 2$  \cite{mpre4}.
\end{figcaption}
\end{center}
\end{figure}

In the rigid spacetime case (found by setting $\Phi_k \equiv 0$), 
the maximum characteristic Floquet exponent
for the $\delta\phi_k$ fluctuation is $\mu_{\rm max}=0.03598$ 
in the sub-Hubble range $3/2<k^2/(\lambda \phi_{\rm co}^2)<\sqrt{3}$, 
where $\phi_{\rm co}$ is the value when the $\phi$ field begins to 
oscillate coherently.  This growth rate is one order smaller than 
that for $\delta\chi_k$.

\begin{figure}
\begin{center}
\singlefig{9cm}{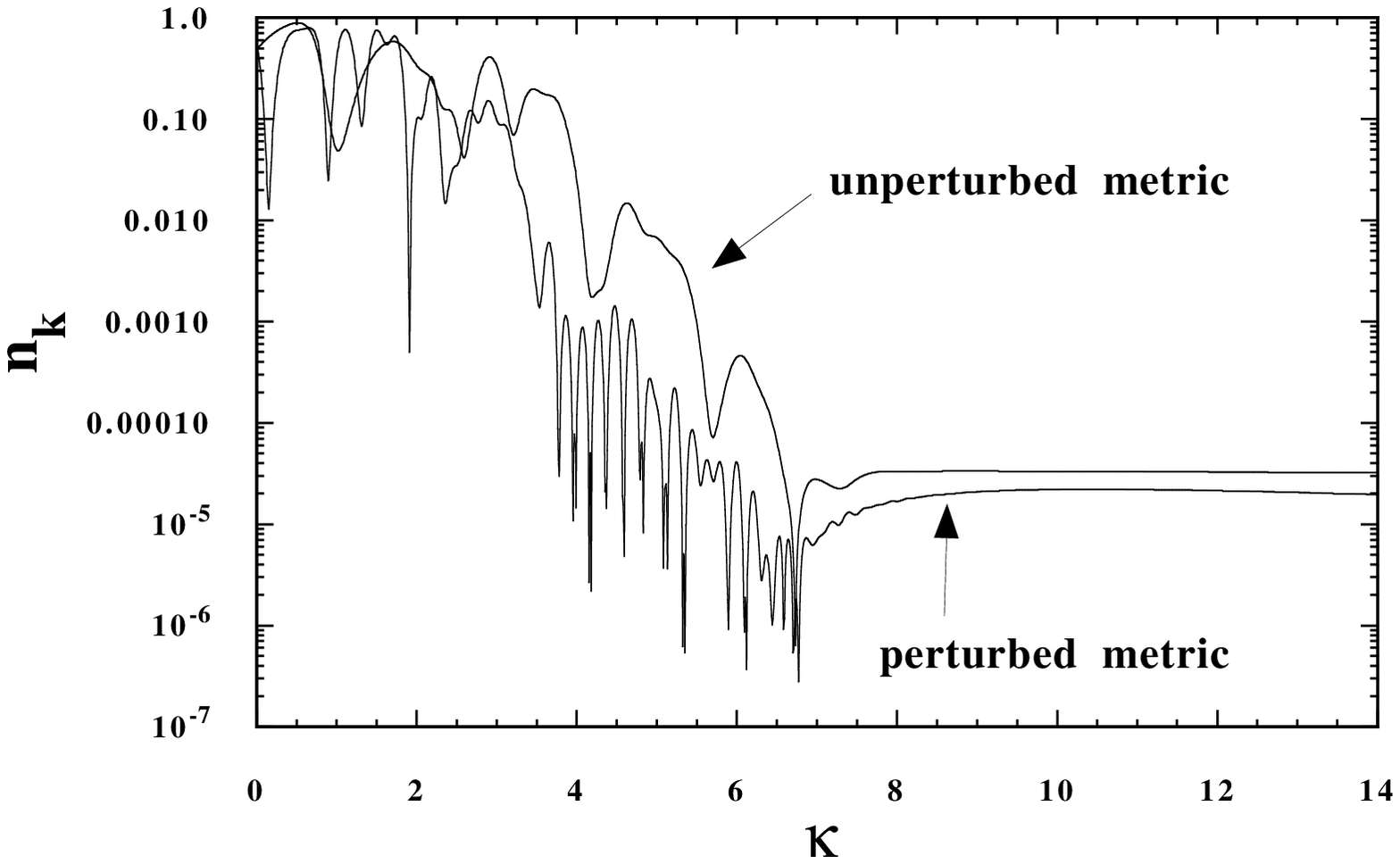}
\begin{figcaption}{Fig6}{9cm}
The spectra of the final comoving number density of fermions $n_k$ 
in the perturbed and unperturbed $(\Phi_k = 0$) metric  
with  $q_1=0$ and $q_2=10^3$   
for $g^2/\lambda=2$ and $\tilde{g}=0$ at $x=50$.
The initial values of scalar fields are chosen as $\phi(0)=0.5m_{\rm pl}$
and $\chi(0)=10^{-9}m_{\rm pl}$. Metric perturbations lead to a decrease in
the number of fermions produced from $\chi$ decays  since they 
enhance $\langle \delta\phi^2 \rangle$  which causes the resonance 
in $\chi$ to end earlier, at smaller values of $\langle\chi^2\rangle$. 
\end{figcaption}
\end{center}
\end{figure}

\begin{figure}
\begin{center}
\singlefig{9cm}{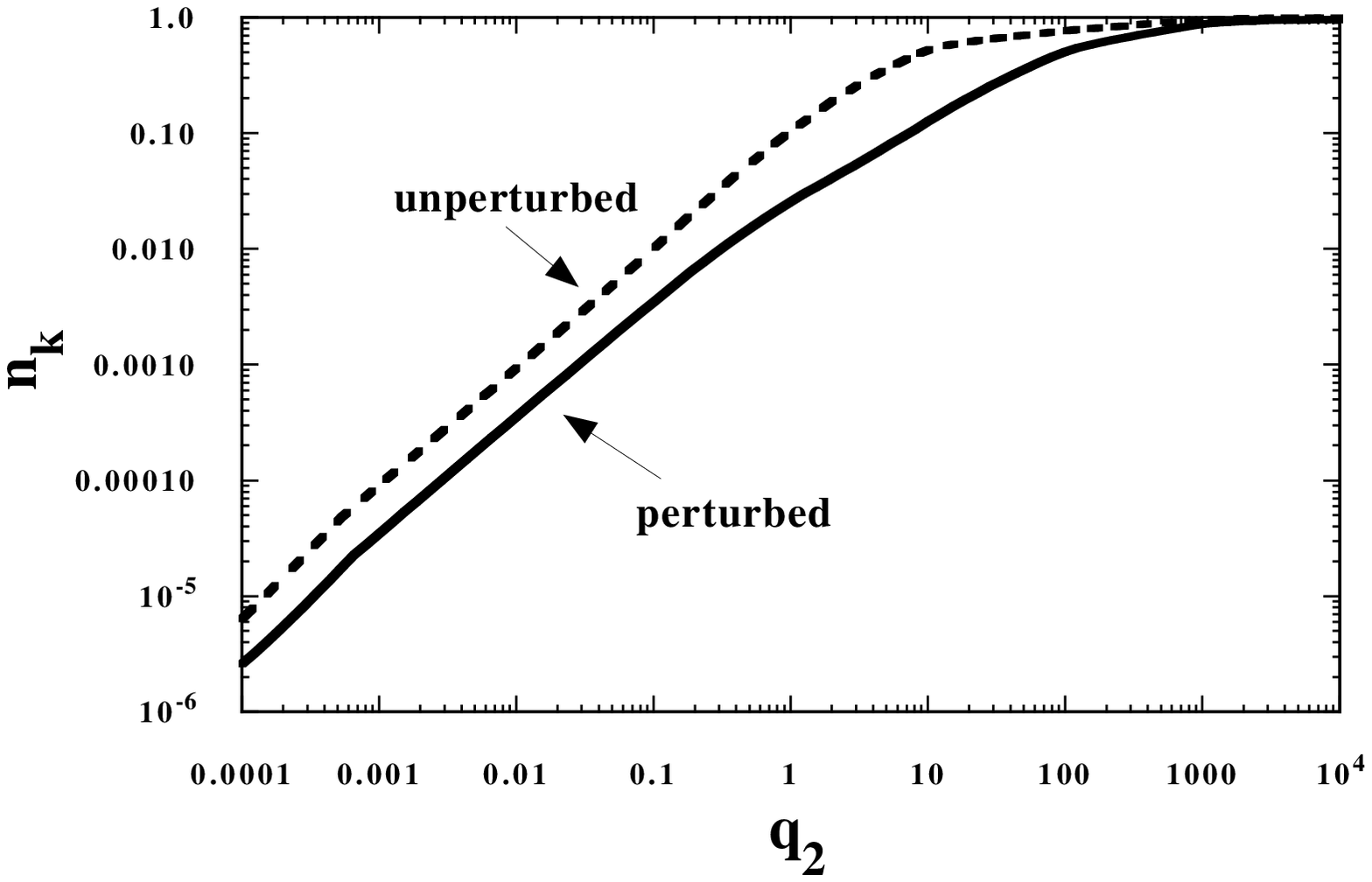}
\begin{figcaption}{Fig7}{9cm}
The maximum comoving number density of fermions $n_k$ with $\kappa = 1$ 
achieved during preheating as a function of $q_2=h_2^2/\lambda$. 
The parameters chosen are $g^2/\lambda = 2, \tilde{g}=0$, and
$q_1 = 0$ with the initial values of $\phi(0)=0.5m_{\rm pl}$
and $\chi(0)=10^{-9}m_{\rm pl}$. The fermion occupation 
number increases almost linearly 
with $q_2$ until it saturates near $n_k = 1$. The effect of the perturbed 
metric $(\Phi_k \neq 0)$ works to reduce the maximum occupation number.
\end{figcaption}
\end{center}
\end{figure}
In the realistic perturbed metric ($\Phi_k \neq 0$), 
however, the rhs of Eq.~$(\ref{B16})$ leads to
the enhancement of the $\delta\phi_k$ fluctuation for super-Hubble modes.
This makes the growth rate of $\langle\delta\phi^2\rangle$ comparable to 
that of $\langle\delta\chi^2\rangle$ as is found in Fig.~\ref{Fig5}.
The final variance $\langle\delta\phi^2\rangle_f \sim 10^{-5}m_{\rm pl}^2$
is about two orders of magnitude larger than in the 
rigid spacetime case in the Hartree approximation \cite{KT1,SB}.

The rapid increase of the $\delta\phi_k$ fluctuation causes 
backreaction to become dynamically important earlier. This means that 
the final variance of 
$\langle\delta\chi^2\rangle$, which we denote $\langle\delta\chi^2\rangle_f$,
is expected to be larger in the rigid case compared with 
the perturbed metric case.
This is verified numerically where we  find that 
$\langle\delta\chi^2\rangle_f$ is larger in the rigid case by about 
one order of magnitude.

Clearly this will have an effect on ``instant'' fermion production due 
to the $h_2$ coupling which  is modified when  metric perturbations are 
taken into account. While the spectra of fermions before  backreaction 
sets in are almost the  same in both the perturbed and unperturbed metric,
including metric perturbations  reduces the maximum occupation 
number of fermions as is evident from Figs.~\ref{Fig6}-\ref{Fig7},
although this effect is rather weak.

This difference practically disappears in the strong coupling regime
$q_2 \gg 1$ due to the saturation of the Pauli bound.
While we find some difference in the weak coupling regime, we warn the 
reader that including second order metric perturbations\cite{ABM} or
rescattering effects\cite{parry} may change these results, 
because this is expected to enhance the homogeneous
$\chi$ field as studied in Ref.~\cite{suppression}.
In addition to this, although we have restricted ourselves 
to the massless inflation, there exist other models of preheating\cite{mpre5} 
where metric perturbations will 
be amplified. 

As long as the $\chi$ field is not strongly suppressed 
during inflation
and gets amplified during preheating, the basic scenario of fermion production
will not be changed.
Nevertheless, in order to understand the effect of metric perturbations on 
fermion production precisely, it is required to study the dynamics of 
fermionic preheating including full backreaction effects 
in broad classes of models, which we leave to the future work.

\subsection{Heavy $\chi$ field: $g^2/\lambda \gg 1$}

In the case $g^2/\lambda \gg 1$, and $\tilde{g}=0$,
the homogeneous $\chi$ and the super-Hubble $\delta\chi_k$ modes 
are exponentially suppressed during inflation since their effective 
masses are much larger than the Hubble 
rate \cite{suppression,suppression2}.
This results in the very small values  $\chi~\lsim~10^{-45}m_{\rm pl}$ 
at the onset of preheating. In this case, fermion production 
by the $\psi$-$\chi$ interaction is very weak, 
although the $\phi$-$\psi$ interaction leads to standard 
fermion production. 

In the presence of the $\tilde{g}^2\phi^3\chi$ coupling, however,
$\langle \chi \rangle \neq 0$ \cite{mpre5}.
In fact, neglecting the derivative terms in Eqs.~$(\ref{B15})$ and
$(\ref{B17})$, one finds the following relations 
at the end of inflation:
\begin{eqnarray}
\chi \approx -\left(\frac{\tilde{g}}{g}\right)^2 \phi,~~~~
\delta\chi_k \approx -3\left(\frac{\tilde{g}}{g}\right)^2
\delta\phi_k.
\label{B23}
\end{eqnarray}
Although these $\chi$ modes are 
about $(\tilde{g}/g)^2$ times smaller than those of the inflaton, they 
undergo parametric amplification during preheating,
which leads to fermion production.
As an example, we depict in Fig.~\ref{Fig8} the spectra 
of the final number density
of fermions for $g^2/\lambda=800$ and $\tilde{g}=10^{-3}g$.
Since this case corresponds to $g^2/\lambda=2l^2$, with 
$l=20$, the $\chi$ modes close to $\kappa=0$ are
strongly enhanced.  In addition to fermion production through the 
$h_1$ coupling, the $\psi$-$\chi$ interaction leads to an extra fermion 
creation as found in Fig.~\ref{Fig8}.

In the case of $g^2/\lambda \gg 1$ with $\tilde{g} > 0$,
the super-Hubble $\delta\chi_k$ fluctuation is about 
$(\tilde{g}/g)^2$ times smaller than the $\delta\phi_k$ fluctuation,
while the sub-Hubble $\delta\chi_k$ is the same order of $\delta\phi_k$.
Since the $\chi$-dependent source term on small scales
in the rhs of Eq.~$(\ref{B18})$ is generally larger than the 
large scale modes for $\tilde{g}~\lsim~g$,
the sub-Hubble metric perturbation grows faster than the the super-Hubble
mode as shown in Fig.~\ref{Fig9}.

\begin{figure}
\begin{center}
\singlefig{9cm}{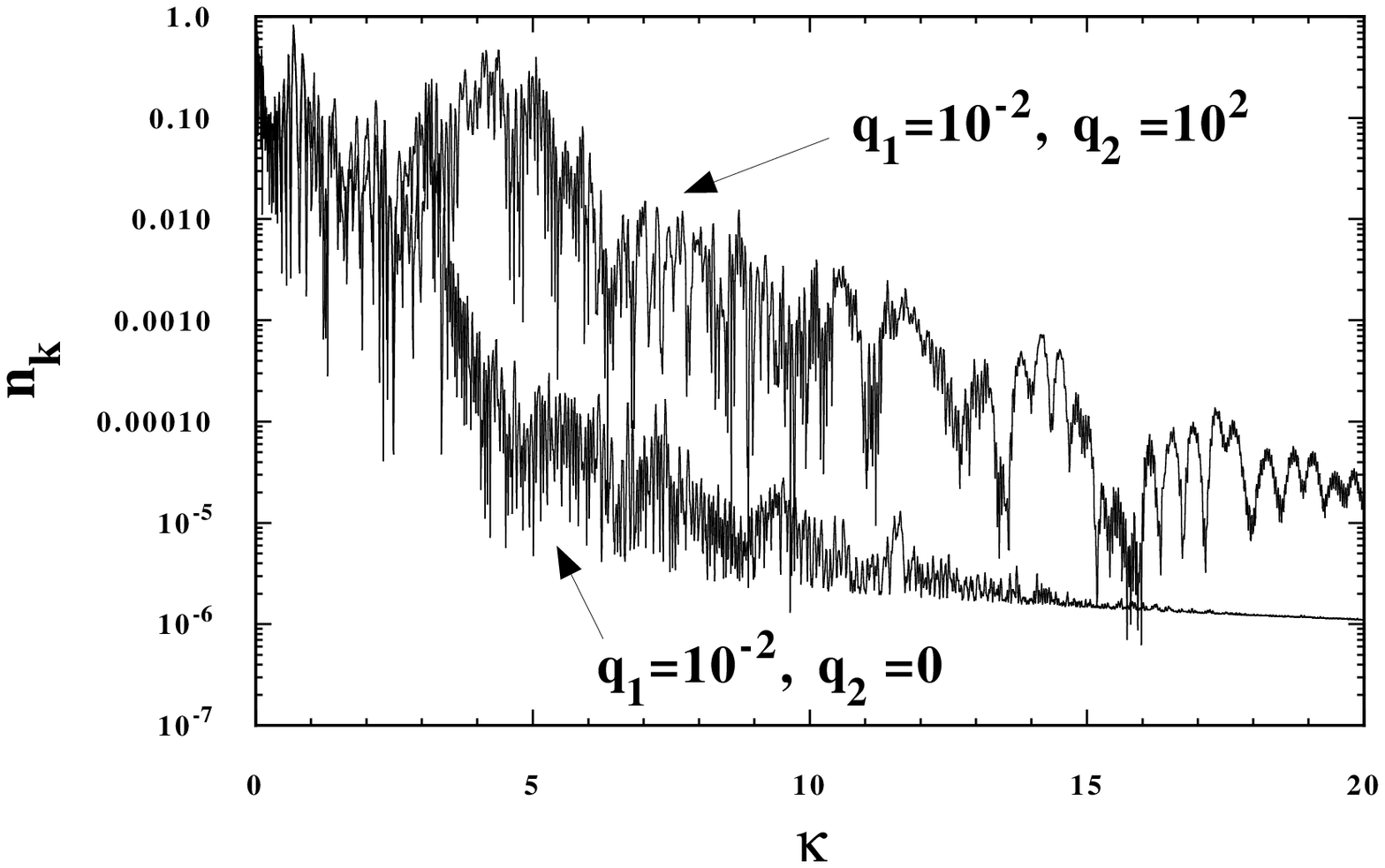}
\begin{figcaption}{Fig8}{9cm}
The final spectra of fermions $n_k$ 
in the cases of $q_1=10^{-2}, q_2=10^2$ (top) and  
$q_1=10^{-2}, q_2=0$ (bottom) with $g^2/\lambda=800$
and $\tilde{g}=10^{-3}g$.
The initial values of scalar fields are chosen as $\phi(0)=0.5m_{\rm pl}$
and $\chi(0)=-(\tilde{g}/g)^2\phi(0)$, values which we use for all figures
in which $\tilde{g} \neq 0$, as dictated by Eq. (\ref{B23})
We find that the existence of the $h_2$ coupling assists  fermion production
as in the case where $g^2/\lambda={\cal O}(1)$.
\end{figcaption}
\end{center}
\end{figure}

For $\tilde{g}~\lsim~10^{-2}g$  amplification of the 
$\delta\chi_k$ fluctuation terminates by backreaction effects
before the super-Hubble metric perturbation begins to grow.
Larger values of $\tilde{g}$ would enhance super-Hubble metric 
perturbations but by requiring the inflaton to have a positive effective mass,
one obtains the following 
relation  making use of Eqs.~(\ref{B14}) and (\ref{B23}) :
\begin{eqnarray}
\frac{\tilde{g}}{g}<\left(\frac{\lambda}{2g^2}\right)^{1/4}.
\label{B24}
\end{eqnarray}
This means that $\tilde{g}/g$ cannot safely exceed unity\footnote{The main
concern is that the inflaton must evolve to the correct vacuum at $\phi = 0$
rather then diverging to infinity.} 
for  $g^2/\lambda \gg 1$, which restricts the strong enhancement of 
super-Hubble metric perturbations, and protects the CMB from preheating.
However, as long as $\tilde{g}/g$ is not much smaller than unity and
the $\chi$ field escapes from the inflationary suppression,
we can say that fermion production is possible only by the $h_2$ coupling
as in the case of $g^2/\lambda={\cal O}(1)$.

\begin{figure}
\begin{center}
\singlefig{9cm}{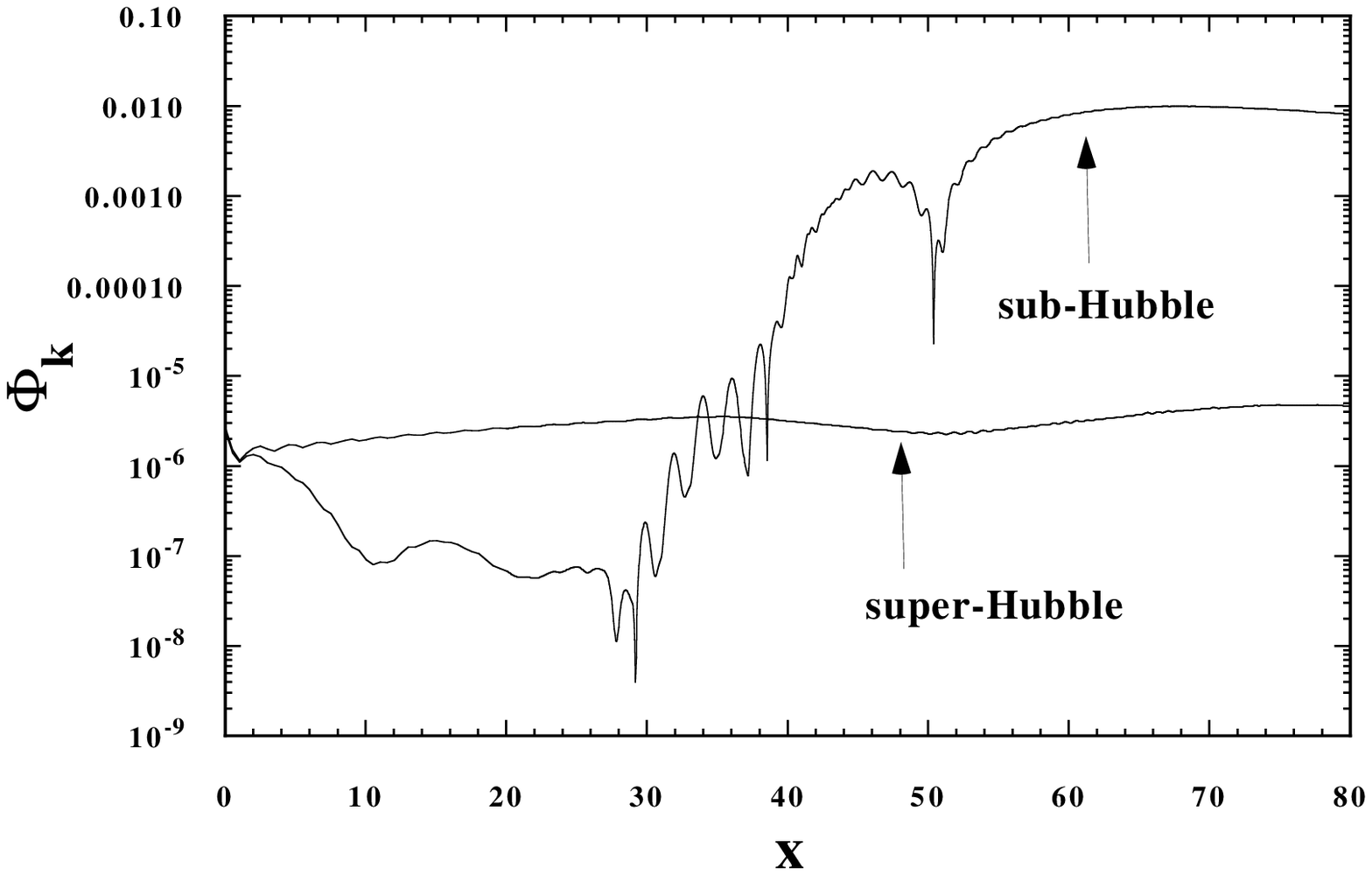}
\begin{figcaption}{Fig9}{9cm}
The evolution of metric perturbations 
$\Phi_k$ for the super-Hubble mode 
$\kappa=10^{-26}$ and the sub-Hubble
mode $\kappa=1$ with $g^2/\lambda=800$
and $\tilde{g}=10^{-3}g$.
The initial values of scalar fields are chosen as $\phi(0)=0.5m_{\rm pl}$
and $\chi(0)=-(\tilde{g}/g)^2\phi(0)$.
In this case, the sub-Hubble metric perturbation gets amplified, while
the super-Hubble mode (and conserved quantity $\zeta$), remain 
almost constant.
\end{figcaption}
\end{center}
\end{figure}

\section{Massive fermion production}\label{massive}

\subsection{No $\chi$ decay: $h_1 \ne 0$, $h_2 = 0$}

If the mass of fermion is taken into account, this generally works  
to suppress efficient fermion production.
Let us first review the case of $h_1 \ne 0$ and $h_2=0$. 
Fermion production takes place around the region where
the effective mass $m_{\rm eff}=m_{\psi}+h_1\phi$ 
vanishes, which corresponds to
\begin{equation}
\phi_*=-m_{\psi}/h_1.
\label{C1}
\end{equation}
As long as the relation $m_{\psi}<|h_1 \phi|$ is satisfied, 
fermions are periodically created when the inflaton passes through $\phi_*$.
After the amplitude of inflaton decreases under the critical value 
$\phi_*$, fermion particle creation ends.

This allows an analytical estimate of the maximum mass of fermions 
produced through the $h_1$ coupling.
Taking note that the conformal 
inflaton field and the scale factor evolve as\cite{structure,SB} 
\begin{equation}
\tilde{\phi}=a\phi \approx \tilde{\phi}_{\rm co}
\cos (0.8472 \sqrt{\lambda}\tilde{\phi}_{\rm co} \eta),
\label{C5}
\end{equation}
\begin{equation}
a \approx \sqrt{\frac{2\pi \lambda}{3}} 
\frac{\tilde{\phi}_{\rm co}^2}{m_{\rm pl}} \eta,
\label{C6}
\end{equation}
the effective mass of the fermion field is approximately 
\begin{equation}
m_{\rm eff}=m_{\psi}+h_1
\sqrt{\frac{3}{2\pi \lambda}} 
\frac{m_{\rm pl}}{\tilde{\phi}_{\rm co}} \frac{1}{\eta}
\cos (0.8472 \sqrt{\lambda}\tilde{\phi}_{\rm co}\eta). 
\label{C7}
\end{equation}
Since the minimum effective mass is achieved for
$0.8472 \sqrt{\lambda}\tilde{\phi}_{\rm co}\eta=\pi$, the upper limit
of the fermion mass production is
\begin{equation}
m_{\psi}~\lsim~\frac{h_1}{5}m_{\rm pl}.
\label{C8}
\end{equation}
This indicates that super-heavy fermions whose masses are of order 
$10^{17}$-$10^{18}$ GeV can be produced if we choose large 
couplings with $h_1$ close to unity \cite{GK2,CKRT}. 

In the vicinity of  $m_{\rm eff}=0$ where particles are
non-adiabatically created, we can approximately express $m_{\rm eff}$ as 
\begin{equation}
m_{\rm eff} = h_1 \phi_*' (\eta-\eta_*),
\label{C2}
\end{equation}
where a prime denotes the derivative with respect to $\eta$.
Since the expansion rate of the universe is typically smaller than 
the creation rate of fermions around $m_{\rm eff}=0$, 
we can safely set $a(\eta_*)=1$, $a'(\eta_*)=0$ in 
Eq.~$(\ref{B45})$. Substituting Eq.~$(\ref{C2})$ for Eq.~$(\ref{B45})$
and introducing new variables 
$\tau \equiv \sqrt{h_1 \phi_*'}(\eta-\eta_*)$, 
$p\equiv k/\sqrt{h_1\phi_*'}$, $d\tau \equiv \sqrt{h_1\phi_*'} d\eta$,
one obtains \cite{CKRT,GK2}
\begin{equation}
\left[\frac{d^2}{d\tau^2}+p^2+\tau^2-i\right]\tilde{\psi}_k=0.
\label{C3}
\end{equation}
The solution of this equation is described by the parabolic cylinder function,
and the comoving number density of fermions is analytically derived 
as\cite{CKRT}
\begin{equation}
n_k \approx \exp\left(-\frac{\pi^2k^2}{\lambda |\phi'_*|}\right)
=\exp\left[-\frac{\pi \phi(0)}{|d\phi_*/dx|}\frac{\kappa^2}
{\sqrt{q_1}}\right].
\label{C4}
\end{equation}
This relation  means that larger values of 
$q_1$ and $|d\phi_*/dx|$  lead to the production of massive 
particles with higher momenta.
Massive fermions can be instantly created even 
after one oscillation of inflaton. Their 
occupation number increases to of order unity for low momentum
modes $\kappa \sim 0$.  Note that
if the coupling $h_1$ satisfies the relation 
$m_{\psi}>|h_1\phi|$ at the beginning of preheating,
the spectrum of Eq.~$(\ref{C4})$ can not be applied.

In order to confirm above analytic estimates, we have numerically 
studied the evolution of the comoving number density 
of fermions. In Fig.~\ref{Fig10}, we plot the evolution of $n_k$ for 
a bare mass $m_{\psi}=10^{-5}m_{\rm pl} \sim 10^{14}$GeV 
with $g^2/\lambda=2$, $\kappa=1$, and $q_1=10^3, 10^4, 10^5$.

In this case, the analytic estimate of Eq.~$(\ref{C8})$ 
predicts that the coupling $h_1$ is required to be 
$h_1~\gsim~5 \times 10^{-5}$ for efficient fermion production,
which corresponds to $q_1~\gsim~2.5\times 10^3$ 
with $\lambda=10^{-12}$. This is elegantly confirmed in
in Fig.~\ref{Fig10} where fermions are not non-adiabatically created 
for the case of $q_1=10^3$. In this case, since the bare mass
$m_{\psi}$ always surpasses the $|h_1\phi|$ term
during the whole stage of preheating, 
we can not expect stochastic increase of $n_k$.

On the other hand, for $q_1=10^4$ and $10^5$,
we find that the comoving number density increases to order 
unity when the inflaton passes through  $\phi_*$.
For  $q_1=10^4$, nonadiabatic amplification of 
fermions occurs only once  around $x \approx 3$, after which 
the amplitude of the inflaton decreases below the critical threshold  
$|\phi_*|$. When $q_1=10^5$, in contrast,  
$n_k$ reaches unity several times. 
If one chooses larger values of $q_1$, nonadiabatic creation 
of fermions takes place periodically for as long as the amplitude of 
inflaton  is larger than $|\phi_*|$.

\begin{figure}
\begin{center}
\singlefig{9cm}{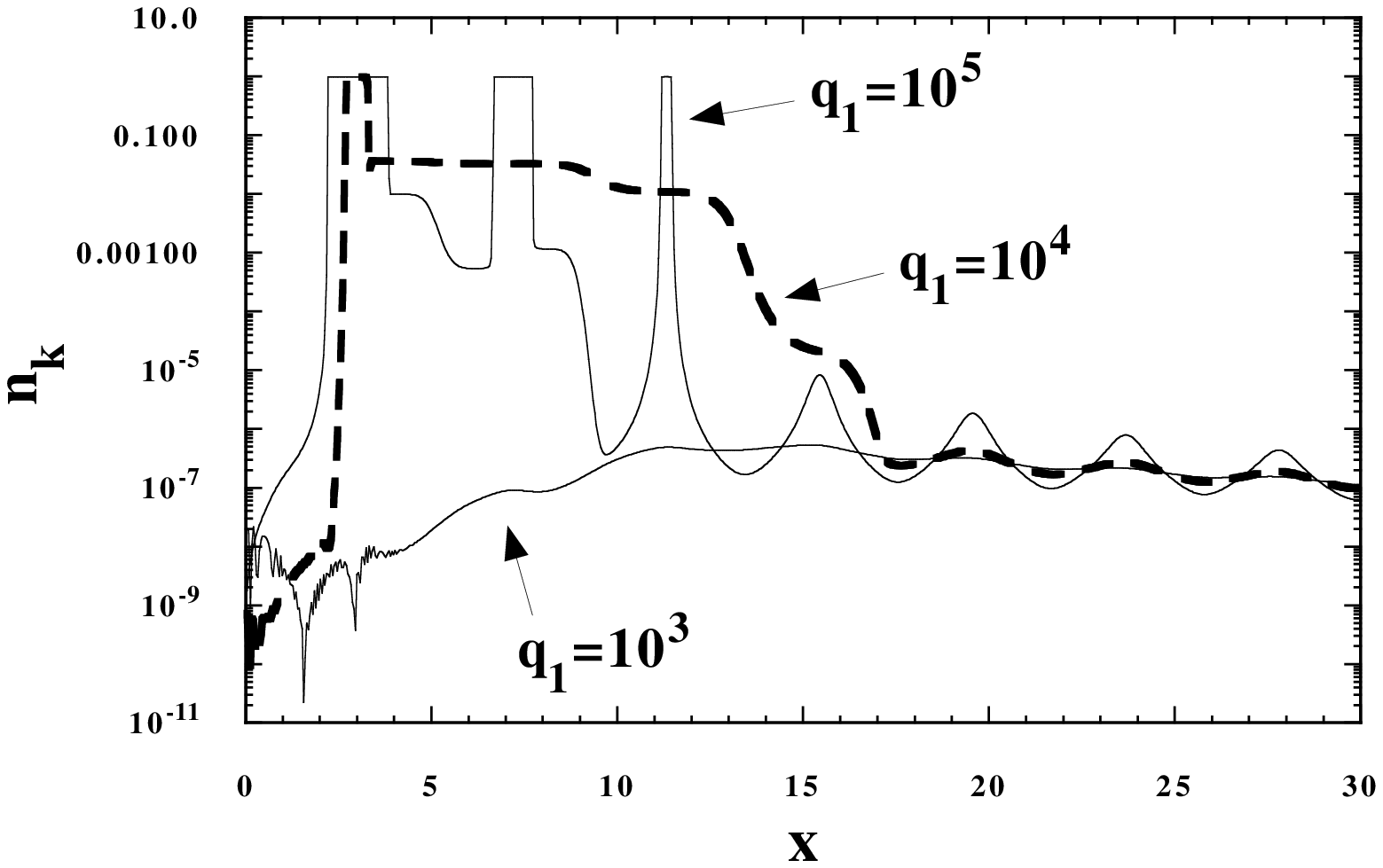}
\begin{figcaption}{Fig10}{9cm}
The time evolution of the comoving number density of fermions $n_k$
with mass $m_{\psi}=10^{-5}m_{\rm pl}$ and the coupling 
$q_1=10^3, 10^4, 10^5$ with $q_2=0$, $g^2/\lambda=2$, and $\tilde{g}=0$
for the sub-Hubble mode $\kappa=1$.
The initial values of the scalar fields are $\phi(0)=0.5m_{\rm pl}$
and $\chi(0)=10^{-9}m_{\rm pl}$.
\end{figcaption}
\end{center}
\end{figure}

In Fig.~\ref{Fig11}, we depict the spectrum  $n_k$ after one oscillation of 
the inflaton with bare fermion mass $m_{\psi}=10^{-5}m_{\rm pl}$ and couplings 
$q_1=10^3, 10^4, 10^5$ with $q_2=0$, $g^2/\lambda=2$, and $\tilde{g}=0$. 
When $q_1=10^4$ and $10^5$ where fermions are created 
around $x \approx 3$, the phase space density of produced particle
is described by Eq.~$(\ref{C4})$. The larger coupling $h_1$ leads to the 
enhancement of higher momentum modes, which we can confirm 
in Fig.~\ref{Fig11}.

Since nonadiabatic fermion production does not occur for 
$q_1~\lsim~2.5\times 10^3$ in this case, the spectrum for $q_1=10^3$
does not obey the analytic estimate of Eq.~$(\ref{C4})$
(see the inset of Fig.~\ref{Fig11}).
Note that the fermion spectra change with the passage of time.
However, for a short period where fermions are created
in the vicinity of $m_{\rm eff}=0$, the spectrum is well described
by Eq.~$(\ref{C4})$.

Super-heavy fermions with masses greater than the GUT scale
$m_{\psi}\sim 10^{16}$ GeV can be produced if the Yukawa-coupling 
is larger than  $h_1 \approx 5 \times 10^{-3}$ by the 
same mechanism described above.
These massive fermions might have  played an important role in leptogenesis 
scenario or be candidates of the super-heavy dark matter\cite{GPRT}.

\begin{figure}
\begin{center}
\singlefig{9cm}{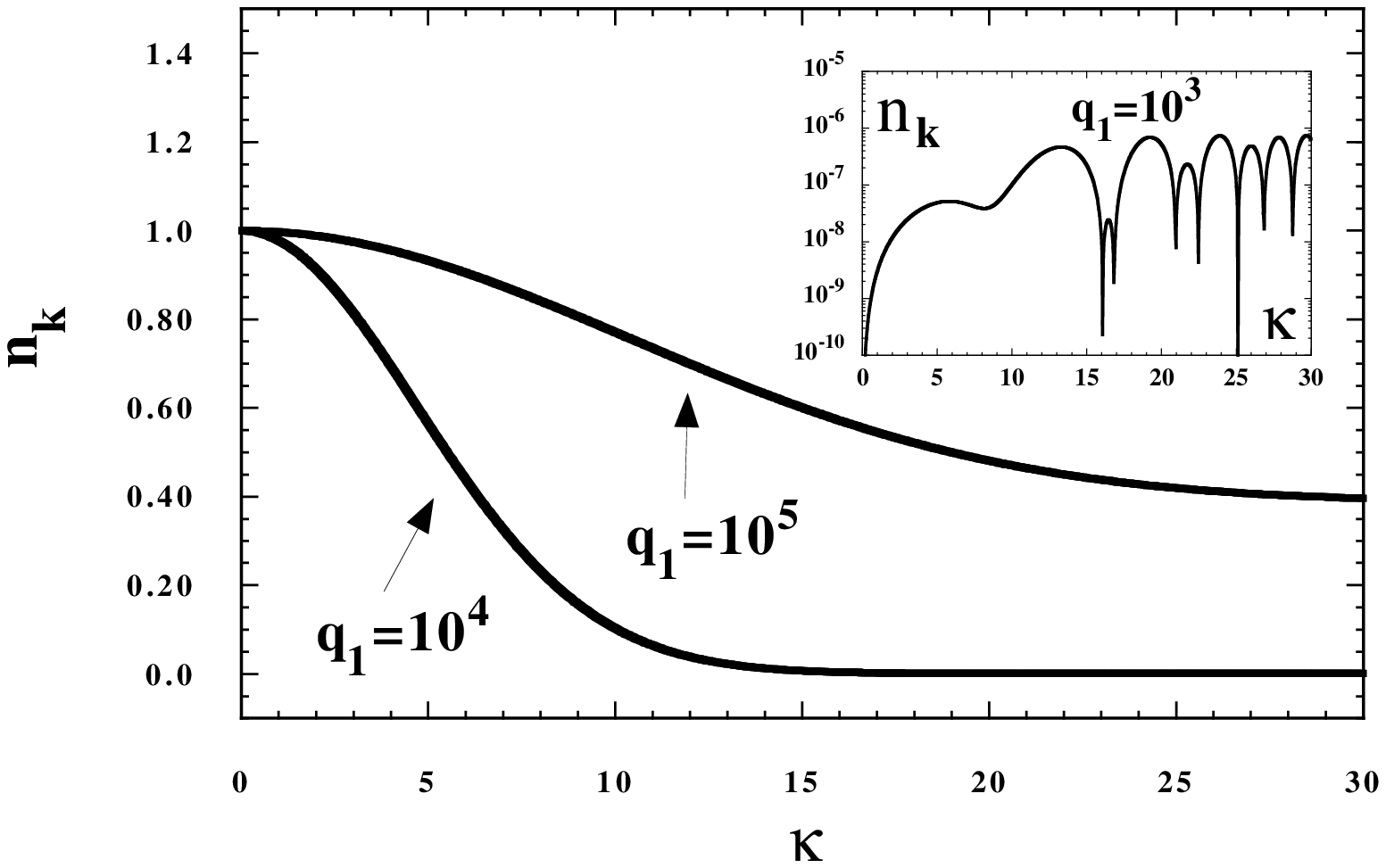}
\begin{figcaption}{Fig11}{9cm}
The spectra of the comoving number density of fermions $n_k$ 
after one oscillation of the inflaton with bare fermion 
mass $m_{\psi}=10^{-5}m_{\rm pl}$ and $g_2 = 0$.
For this mass, simple analytical estimates imply a threshold for instant 
production at around $q_1 = 2.5 \times 10^3$. The main figure shows $n_k$
for  $q_1=10^4, 10^5$ with $g^2/\lambda=2$, and 
$\tilde{g}=0$. The spectra match the simple analytical result of 
Eq. (\ref{C4}) very well. 
{\bf Inset}: $n_k$ vs $\kappa$ for $q_1$ below the threshold, $q_1=10^3$. 
Essentially no instant production of fermions takes place. 
The initial values of scalar fields are chosen to be $\phi(0)=0.5m_{\rm pl}$
and $\chi(0)=10^{-9}m_{\rm pl}$.
\end{figcaption}
\end{center}
\end{figure}

\subsection{Instant preheating considered: $h_1=0$, $h_2 \neq 0$}

Let us next consider the case with $h_1=0$ and $h_2 \ne 0$.
When $1<g^2/\lambda<3$, since long-wavelength modes of the 
$\chi$ field are resonantly
amplified, one might assume that massive fermions are
efficiently produced from the beginning of preheating.
However, for instant production of super-massive fermions,
the whole $h_2\chi$ term in Eq.~(\ref{effmass}) must  be 
comparable to the bare mass $m_{\psi}$ at the start of preheating.

This condition is rather difficult to achieve unless both the coupling 
$h_2$ and the initial value of $\chi$ are large.
For example, for an initial value $\chi(0)=10^{-9}m_{\rm pl}$
which we used in the simulations of Figs.~\ref{Fig2}-\ref{Fig7}, 
even strong couplings  $h_2 \sim 1$ yield 
$h_2\chi(0) \sim 10^{-9}m_{\rm pl}$.
This indicates that super-massive fermions heavier than
$m_{\psi} \sim 10^{-6}m_{\rm pl}$ cannot be
produced at the initial stage.
In this case however, once the amplitude of the $\chi$ field grows 
through parametric resonance and the $|h_2\chi|$ becomes larger than 
$m_{\psi}$, $\chi-\psi$ production begins to be
relevant.

If $g^2/\lambda={\cal O}(1)$ and $\tilde{g} = 0$,  
large initial $\chi$ values 
$\chi(0) \gg 10^{-9}m_{\rm pl}$ will lead to the strong 
amplification of super-Hubble metric perturbations. 
In order that the model not conflict with observations one requires 
$\chi(0) < 10^{-9}m_{\rm pl}$ which means that 
instant fermion production of massive fermions is rather difficult 
unless $h_2$ is unnaturally large.

When $g^2/\lambda \gg 1$ with the $\tilde{g}$ 
coupling, the $\chi$ field is constrained by Eq.~$(\ref{B23})$.
Since the super-Hubble $\delta\chi_k$ fluctuation is about 
$(\tilde{g}/g)^2$ times smaller relative to $\delta\phi_k$,
even the initial value of $\chi(0) \sim  10^{-6}m_{\rm pl}$
does not result in the strong enhancement of super-Hubble
metric perturbations as shown in Fig.~\ref{Fig9}.
Making use of the relation $(\ref{B23})$, 
the effective mass of fermions at the end of 
inflation is approximately given by
\begin{equation}
m_{\rm eff} \approx m_{\psi}-h_2\left(\frac{\tilde{g}}{g}
\right)^2 \phi(0).
\label{C9}
\end{equation}
When $h_2$ and $\phi(0)$ are positive, 
$m_{\rm eff}$ can take negative values 
at the end of inflation for large couplings $h_2$ and
ratio $\tilde{g}/g$. 
Since the amplitude of the $\chi$ field decreases 
at the initial stage of preheating due to cosmic expansion,
instant fermion production
takes place when $m_{\rm eff}$ vanishes before 
the $\chi$ fluctuation is amplified through resonance. 
This case corresponds to
\begin{equation}
h_2\left(\frac{\tilde{g}}{g}\right)^2~\gsim~
2\frac{m_{\psi}}{m_{\rm pl}},
\label{C10}
\end{equation}
where we used the value $\phi(0)=0.5m_{\rm pl}$. 

For example, for a coupling of $h_2=1$ and 
$\tilde{g}/g=10^{-2}$,  fermions with bare masses
of order $m_{\psi}\sim 5 \times 10^{-5}m_{\rm pl}\sim
5 \times 10^{14}$ GeV are produced at the initial stage.
For instant creation of  GUT scale fermions 
$m_{\psi}\sim 10^{16}$ GeV, 
we require $h_2(\tilde{g}/g)^2~\gsim~10^{-3}$.
Due to the existence of the $\tilde{g}/g$ term, which is required to be
smaller than unity for successful 
inflation, massive fermion production via the $h_2$
coupling is generally hard to realize relative to the $h_1$ 
case [c.f. Eq.~$(\ref{C8})]$.

It is important to recognise that Eq.~$(\ref{C10})$ is
the condition of instant fermion production at the initial stage 
of preheating.
Even when $h_2\chi(0) \ll m_{\psi}$ initially, subsequent exponential growth 
of $\chi$ leads to fermion production at the later stages of preheating. 

These effects are illustrated in Fig.~\ref{Fig12}, where we 
show the evolution of $n_k$ for fermions with mass 
$m_{\psi}=10^{-6}m_{\rm pl}$ and coupling 
$g^2/\lambda=5.0\times 10^3$, $\tilde{g}/g=3.0 \times 10^{-3}$,
$h_2=1$ and $0.1$ for the mode $\kappa=2$.
The condition of Eq.~$(\ref{C10})$ is satisfied for $h_2=1$,
which leads to instant fermion production  found
in Fig.~\ref{Fig12}. Since the expansion of the universe reduces the 
amplitude of the $\chi$ field at the first stage of preheating,
fermionic preheating is frozen for some time
once $m_{\rm eff}$ vanishes at $x\sim 2$. For low momentum modes
$\kappa \sim 0$, $n_k$ increases to unity by instant fermion production.

For $x~\gsim~30$, parametric amplification of 
the $\chi$ field causes nonadiabatic fermion production
(see Fig.~\ref{Fig12}).
For $h_2=0.1$, where the condition of Eq.~$(\ref{C10})$
is not satisfied, $n_k$ does not increase at the initial stage.
However, $n_k$ begins to grow after the $\chi$ field has been  
sufficiently amplified.
Since the homogeneous $\chi$ field periodically changes between 
positive and negative values, we find that fermion productions occurs
in the same way even for the case of $h_2<0$ at the final stage of preheating.

\begin{figure}
\begin{center}
\singlefig{9cm}{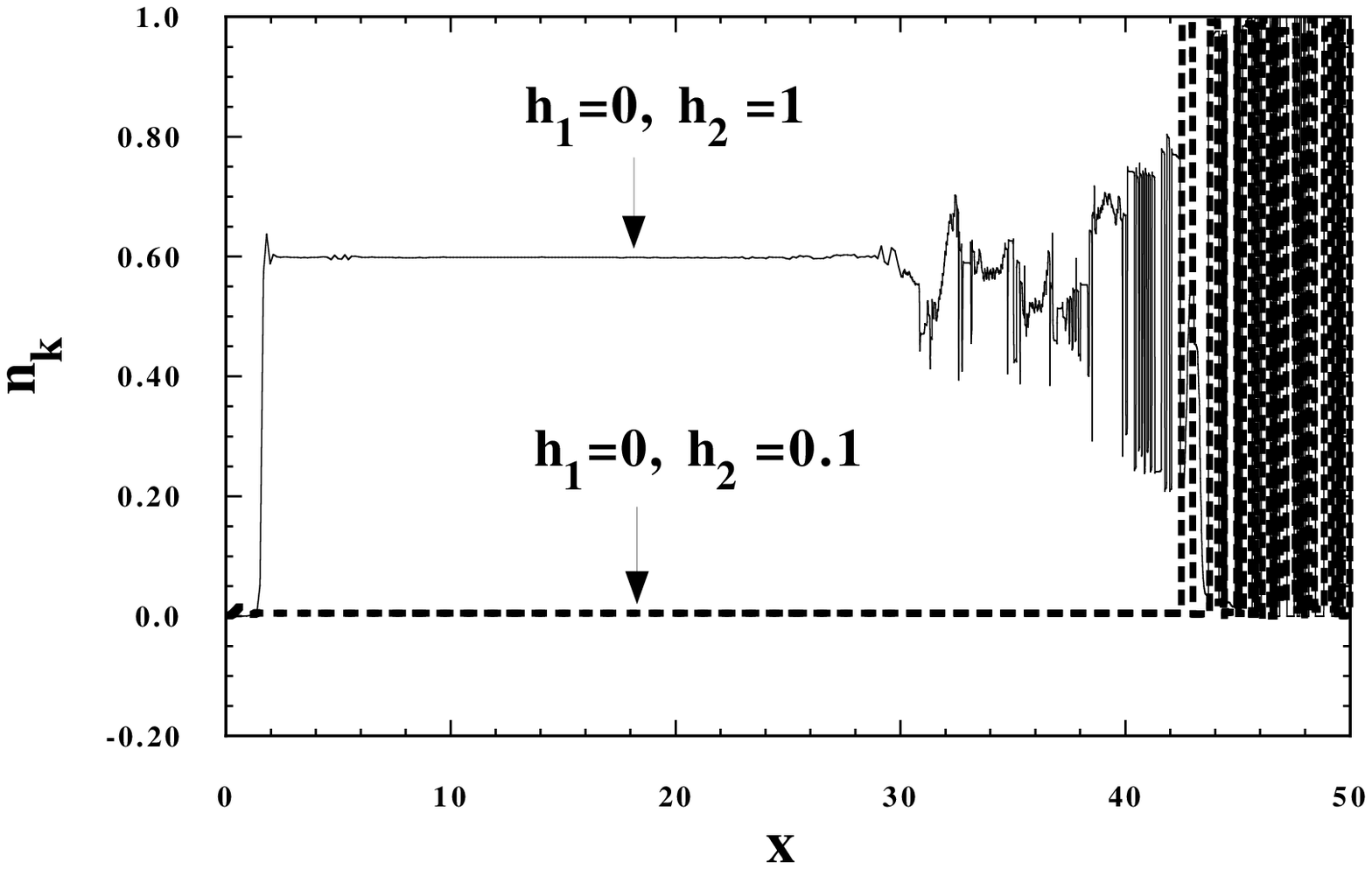}
\begin{figcaption}{Fig12}{9cm}
Instant production followed by chaotic production of fermions. 
The evolution of the comoving number density of fermions $n_k$ with bare mass 
$m_{\psi}=10^{-6}m_{\rm pl}$ and the coupling $h_2=1$ (solid), 
$h_2=0.1$ (dotted) with $h_1=0$, $g^2/\lambda=5.0\times 10^3$, 
and $\tilde{g}/g=3.0 \times 10^{-3}$ for the mode $\kappa=2$.
The initial conditions are  $\phi(0)=0.5m_{\rm pl}$
and $\chi(0)=-(\tilde{g}/g)^2\phi(0)$.
\end{figcaption}
\end{center}
\end{figure}

\begin{figure}
\begin{center}
\singlefig{9cm}{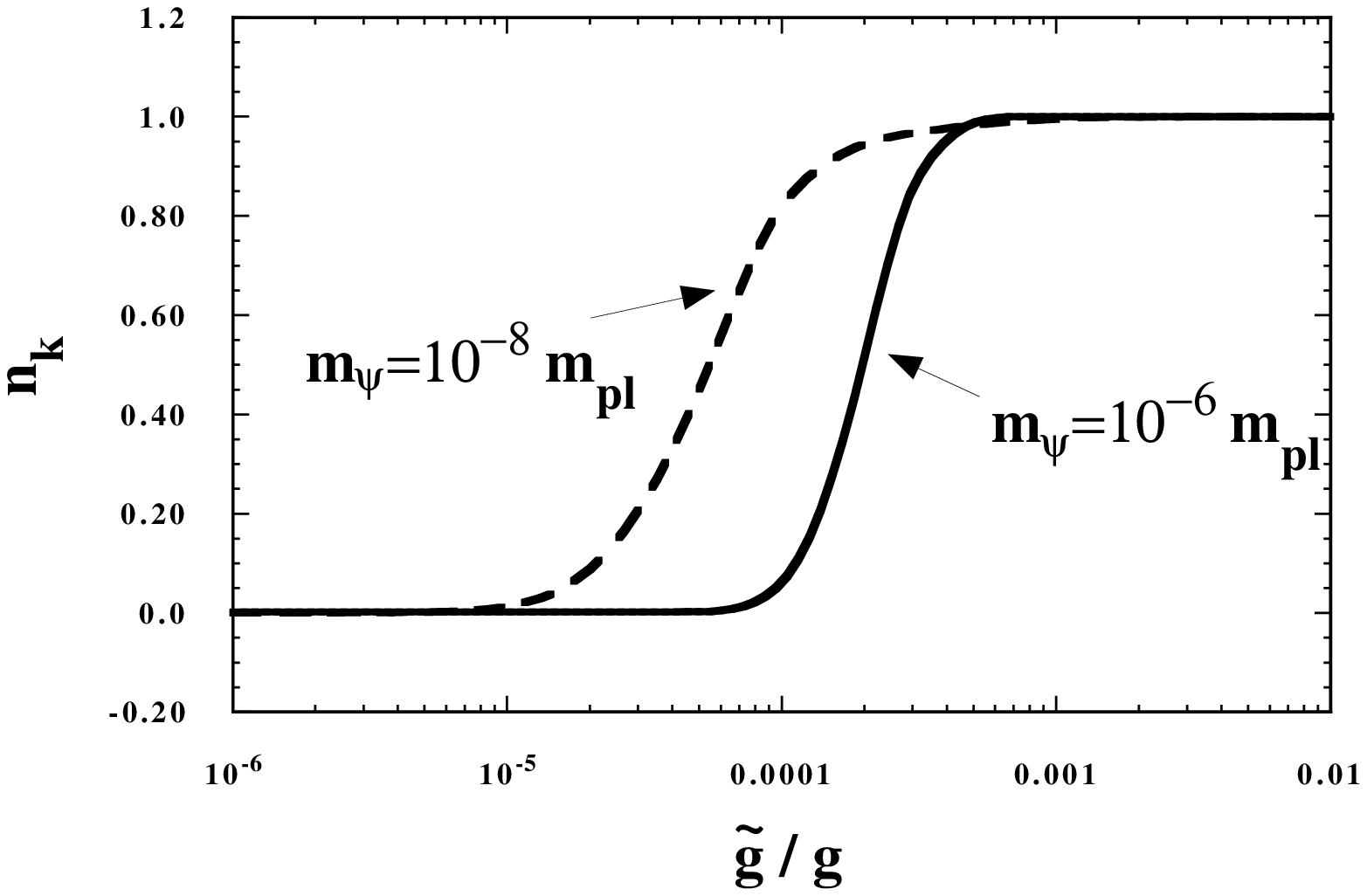}
\begin{figcaption}{Fig13}{9cm}
The maximum comoving number density of fermions $n_k$ for the Hubble-scale
mode $\kappa = 1$ as a function of $\tilde{g}/g$ for 
$g^2/\lambda = 5.0 \times 10^3$, 
$h_1 =0$, and $h_2 =1$ with mass $m_{\psi}=10^{-6}m_{\rm pl}$
and $m_{\psi}=10^{-8}m_{\rm pl}$. As the fermion mass increases, large 
$\tilde{g}/g$ is required before $\chi \rightarrow \psi$ decays 
become efficient. 
\end{figcaption}
\end{center}
\end{figure}
In Fig.~\ref{Fig13}, we show the maximum number density of fermions as a 
function of $\tilde{g}/g$ in two cases of $m_{\psi}=10^{-6}m_{\rm pl}$
and $m_{\psi}=10^{-8}m_{\rm pl}$ with $g^2/\lambda = 5 \times 10^3$, 
$h_1 =0$, and $h_2 =1$. We find that large values of  $\tilde{g}/g$  are 
needed as one increases the fermion bare mass. In order to produce GUT 
scale fermions $m_{\psi} \sim 10^{16}$ GeV, we require strong coupling 
$\tilde{g}/g$ close to its upper bound of Eq.~$(\ref{B24})$.

\subsection{Both couplings taken into account: $h_1 \ne 0$, $h_2 \ne 0$}

If both $h_1$ and $h_2$ are non-zero, 
fermionic preheating strongly depends both upon the absolute and relative 
strengths of  the couplings. In the presence of the $\tilde{g}$ coupling, 
the effective mass of fermion at the initial stage of preheating is 
approximately given by 
\begin{equation}
m_{\rm eff} \approx m_{\psi}+\left[ h_1-h_2
\left(\frac{\tilde{g}}{g}\right)^2 \right]\phi.
\label{C11}
\end{equation}
This relation indicates that the negative coupling $h_2$ will assist 
fermion production caused by the positive $h_1$ coupling.
Conversely, when $h_1$ and $h_2$ are both positive, including
the $h_2$ coupling generally works as to {\em suppress} 
fermion production as long as $h_1 \ge h_2(\tilde{g}/g)^2$. 

Let us consider concrete cases for the couplings:  
$h_1=10^{-5}$, $h_2=1$, $g^2/\lambda=5.0 \times 10^3$, and 
$\tilde{g}/g=3.0 \times 10^{-3}$
with  $m_{\psi}=5.0 \times 10^{-7}m_{\rm pl}$. 
When $h_2 = 0$, fermions are stochastically
produced at the initial phase of preheating  when the inflaton passes 
causes $m_{\rm eff}$ to vanish.

\begin{figure}
\begin{center}
\singlefig{9cm}{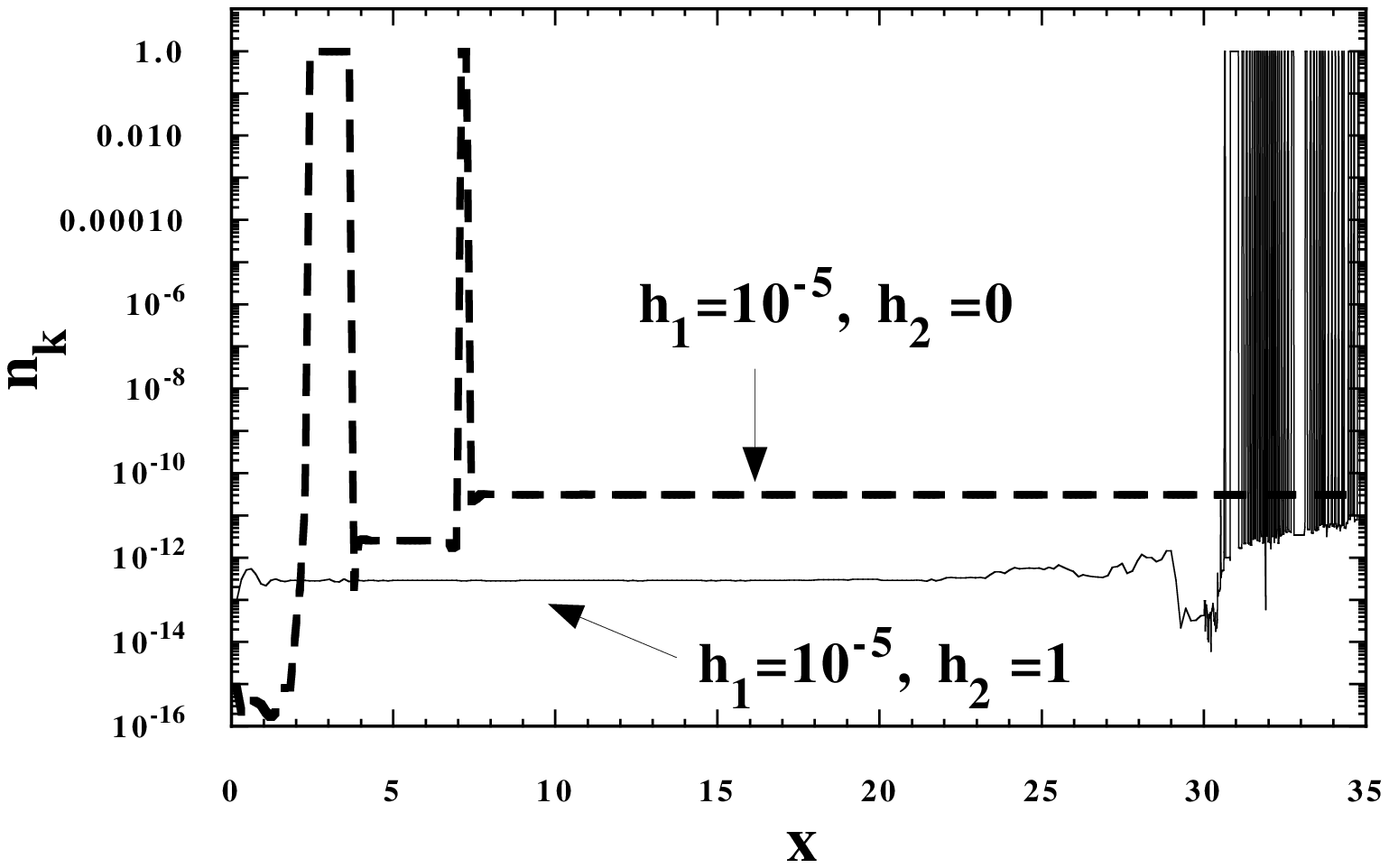}
\begin{figcaption}{Fig14}{9cm}
$h_2$-suppression of instant fermion production.
Number density of fermions $n_k$ vs $x$ in the cases 
$h_1=10^{-5}$, $h_2=1$  (solid), and $h_1=10^{-5}$, $h_2=0$ (dotted),
with $g^2/\lambda=5.0\times 10^3$, $\tilde{g}/g=3.0 \times 10^{-3}$, and 
$m_{\psi}=5.0 \times 10^{-7}m_{\rm pl}$ for the mode $\kappa=1$.
The initial conditions chosen were  $\phi(0)=0.5m_{\rm pl}$
and $\chi(0)=-(\tilde{g}/g)^2\phi(0)$. 
The $h_2$ coupling reduces the violation of the adiabatic condition 
and removes the initial instant production present when $h_2 = 0$ for an
otherwise equivalent  parameter set. 
\end{figcaption}
\end{center}
\end{figure}
However, the $h_2$ coupling makes the second term in 
Eq.~$(\ref{C11})$ smaller than the bare mass $m_{\psi}$.
Hence instant fermion production 
does not take place in this case, although $n_k$ increases 
for $x~\gsim~30$ due to the resonance from the $h_2$ coupling
(see Fig.~\ref{Fig14}).

This suppression is relevant for large coupling $h_2$ relative 
to $h_1$. When $h_1$ and $h_2$ are of the same order, fermion
production mainly proceeds by the $h_1$ coupling, because 
the ratio $\tilde{g}/g$ is typically smaller than unity.
 
When $h_1=0$ and $h_2 \ne 0$, massive fermions which satisfy the 
relation $(\ref{C10})$ are instantly produced. This typically 
requires rather large coupling $h_2$ for producing
massive fermions whose masses are greater than 
$m_{\psi}\sim 10^{-6}m_{\rm pl}$.
However, introducing the $h_1$ coupling even much smaller than 
$h_2$ can potentially alter the dynamics of fermion production. 

For example, in the case of  
$h_1=0$, $h_2=1$, $g^2/\lambda=5.0 \times 10^3$, and 
$\tilde{g}/g=3.0 \times 10^{-3}$ with  $m_{\psi}=10^{-5}m_{\rm pl}$,
we can not expect stochastic
amplification of fermions at the initial stage while $n_k$ increases for
$x~\gsim~40$ after the $\chi$ field is sufficiently produced.
Including the $h_1$ coupling greater than $10^{-4}$, fermions are 
periodically created from the beginning of preheating (see Fig.~\ref{Fig15}).
This means that the instant preheating scenario discussed in 
Ref.~\cite{instant} can be further strengthened by taking into account 
the $h_1$ coupling.

\begin{figure}
\begin{center}
\singlefig{9cm}{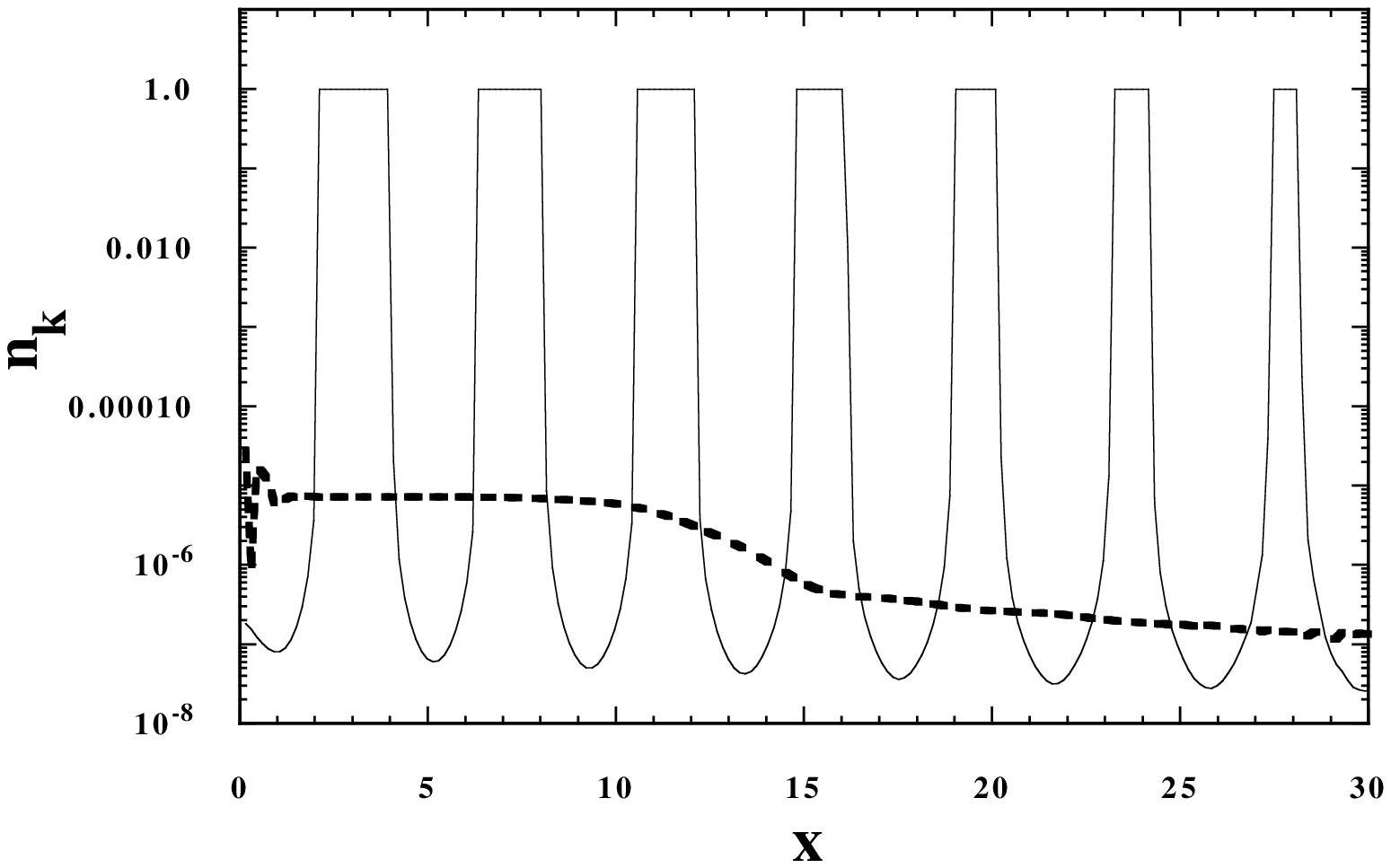}
\begin{figcaption}{Fig15}{9cm}
Significant instant production. 
The time evolution of  fermion number density  $n_k$ 
for $h_1=10^{-3}$, $h_2=1$ (solid) and $h_1=0$, $h_2=1$ (dotted)
with $g^2/\lambda=5.0\times 10^3$, $\tilde{g}/g=3.0 \times 10^{-3}$, and 
$m_{\psi}=10^{-5}m_{\rm pl}$ for the mode $\kappa=1$.
The initial scalar field amplitudes are  $\phi(0)=0.5m_{\rm pl}$
and $\chi(0)=-(\tilde{g}/g)^2\phi(0)$.
\end{figcaption}
\end{center}
\end{figure}

\section{Concluding remarks and discussion}   

Production of gravitinos and other fermions  after inflation  has become 
an issue of great importance in modern cosmology since their creation
provides a window into the early universe and provides a powerful 
mechanism for distinguishing between different models and 
regions of parameter space.
As such the phenomenon of fermion creation deserves to be studied in 
great detail, especially as unusual and  non-trivial results are 
common as the complexity of the model scenario  is increased. 

With this motivation  we have studied 
spin-$1/2$ fermion production over a $6$-dimensional 
coupling constant parameter space with both and without 
metric perturbations included. 
This parameter space spans both the direct resonant decay of the 
inflaton into both massive and massless fermions and the indirect 
$\phi \rightarrow \chi \rightarrow \psi$ decays of instant preheating. 
These decays are relevant since they model the couplings expected for the 
gravitino (through the K\"ahler potential) in realistic supergravity 
theories.    

Previous studies have limited themselves to $3$ or $4$-dimensional 
parameter spaces in which either there is only the $\phi$-$\psi$ system 
\cite{BHP,RHS,GK} or in which the $\chi$ field has zero vacuum 
expectation value during inflation and there is no $\phi$-$\psi$ 
coupling \cite{instant}.

We have verified the general expectation that increasing 
the $\phi$-$\psi$ ($h_1$) and 
$\chi$-$\psi$ ($h_2$) couplings typically leads to enhanced fermion 
production though with an interesting exception. In the 
case where $\langle \chi \rangle > 0$ and 
both $h_1$ and $h_2$ are  non-negative, increasing $h_2$ leads to a 
{\em decrease}  in the production of fermions due to a weakening of the
degree of  violation of the adiabatic condition, c.f. Eq. (\ref{C11}).

Further, if $\langle \chi \rangle = 0$ during inflation 
(i.e. $\tilde{g} = 0$) instant preheating is initially sub-dominant relative
to direct decays of the inflaton (Fig.~\ref{Fig3}) but later becomes important 
due to resonant growth of the $\chi$ field. If $\langle \chi \rangle \neq 0$
during inflation an initial burst of instant preheating is typically followed
by a quiescent period and then a sudden chaotic burst of fermion production 
when the $\chi$ field becomes large due to resonance
 (see Figs.~\ref{Fig12},\ref{Fig14}). 

In addition we found that metric perturbations had an 
important impact on $n_k$ at weak coupling (small $h_1, h_2$) 
while their effect is weak  at strong coupling due to the saturation
caused by the Pauli exclusion principle. In the cases we studied 
(Figs.~\ref{Fig6}, \ref{Fig7}) the effect of metric perturbations was to 
{\em reduce} the production of fermions. 
This can be understood as due to the enhancement of 
the inflaton resonance and hence inflaton variance 
$\langle\delta\phi^2\rangle$ when metric perturbations
are included\cite{mpre2b,mpre4}. 
This causes backreaction to end the resonance earlier,
before the inflaton and $\chi$ fields have decayed significantly to  
fermions. 

Fermion production via the $h_2$ coupling is sensitive to the value of 
$\chi$ at the end of inflation. Although we considered the massless 
inflaton model where the $\chi$ field may escape  suppression during 
inflation for $g^2/\lambda={\cal O}(1)$, there exists other scenarios 
of inflation in which the exponential suppression can be avoided.
One such scenario is a nonminimal coupled $\chi$ 
field\cite{nonminimalpre,mpre4}.
During inflation, the homogeneous and super-Hubble $\chi$ modes exhibit 
exponential growth for negative nonminimal coupling $\xi$
as pointed out in Ref.~\cite{SH}, which may drastically change 
the scenario of instant fermionic preheating described in this paper. 
This amplification of the $\chi$ field is expected to work even 
for small values
of $|\xi|$  smaller than unity in a model-independent way. 
Nevertheless this requires a detailed study including the 
evolution of metric perturbations
during inflation and preheating, which we leave to the future work.

\section*{ACKNOWLEDGEMENTS}
We thank J\"urgen Baacke, Mar Bastero-Gil, Christopher Gordon, 
David Kaiser, Kei-ichi Maeda, and Takashi Torii for useful discussions.
ST was supported partially by
a Grant-in-Aid for Scientific Research Fund of the Ministry of
Education, Science and Culture (No. 09410217 and Specially
Promoted Research No. 08102010), by the Waseda University
Grant for Special Research Projects. FV is supported by CONYCIT scholarship 
Ref:115625/116522.


\end{document}